\documentclass[journal=jacsat,manuscript=article]{achemso}

\usepackage{amsmath,amssymb,amsfonts}%
\usepackage{amsthm}%
\usepackage{mathrsfs}%
\usepackage{xcolor}
\usepackage{tikz}
\usepackage{booktabs}
\usepackage{subcaption}
\usepackage{multirow}
\usepackage[version=3]{mhchem} % Formula subscripts using \ce{}
\expandafter\def\csname SIref@sec:slab_tests\endcsname{S1}
\expandafter\def\csname SIref@fig:7A_vs_14A_mlff\endcsname{S1}
\expandafter\def\csname SIref@tab:7A_vs_14A_dft\endcsname{S1}
\expandafter\def\csname SIref@fig_predictions\endcsname{S2}
\expandafter\def\csname SIref@tab:wmat\endcsname{S2}
\expandafter\def\csname SIref@tab:rmat\endcsname{S3}
\expandafter\def\csname SIref@sec:anomalies\endcsname{S2}
\expandafter\def\csname SIref@fig:anomaly_bins\endcsname{S3}
\expandafter\def\csname SIref@fig_full_clustering\endcsname{S4}
\expandafter\def\csname SIref@fig_closest_noCuZn\endcsname{S5}
\expandafter\def\csname SIref@fig_closest_Ni110\endcsname{S6}
\expandafter\def\csname SIref@fig:aed_groups\endcsname{S7}
\expandafter\def\csname SIref@fig_umap\endcsname{S8}
\expandafter\def\csname SIref@fig:relative_contri_pca\endcsname{S9}
\expandafter\def\csname SIref@fig:ternary_mixture\endcsname{S10}
\expandafter\def\csname SIref@fig_sensitivity_analysis\endcsname{S11}
\expandafter\def\csname SIref@fig:ads_energy_ranges\endcsname{S12}
\expandafter\def\csname SIref@tab:selected_facets\endcsname{S4}
\expandafter\def\csname SIref@tab:selected_facets_Ni110\endcsname{S5}
\expandafter\def\csname SIref@fig:sp-nsp\endcsname{S13}
\expandafter\def\csname SIref@tab:sp-nsp\endcsname{S6}
\expandafter\def\csname SIref@sec:cu4pd_casestudy\endcsname{S3}
\expandafter\def\csname SIref@fig:cu4pd_pca\endcsname{S14}
\expandafter\def\csname SIref@tab:case_study\endcsname{S7}

\newcommand*\SIref[1]{\expandafter\ifx\csname SIref@#1\endcsname\relax\textbf{S??}\else\csname SIref@#1\endcsname\fi}

\author{Prajwal Pisal}
\affiliation[Technical University of Munich]
{Department of Physics, Technical University of Munich,
James-Franck-Strasse 1, 85748 Garching,  Germany.}
\alsoaffiliation[Aalto University]
{Department of Applied Physics, Aalto University, P.O. Box 11000, AALTO, FI-00076, Finland.}
\alsoaffiliation[MCML]
{Munich Center for Machine Learning, Germany.}
\alsoaffiliation[AMC]
{Atomistic Modeling Center, Munich Data Science Institute, Technical University of Munich, Walther-von-Dyck Str. 10, 85748 Garching, Germany}
\author{Ondrej Krejci}
\affiliation[CHEM - Aalto University]
{Department of Chemistry and Materials Science, Aalto University, P.O. Box 11000, AALTO, FI-00076, Finland.}
\alsoaffiliation[Aalto University]
{Department of Applied Physics, Aalto University, P.O. Box 11000, AALTO, FI-00076, Finland.}
\alsoaffiliation{Department of Mechanical and Materials Engineering, University of
Turku, Vesilinnantie 5, Turku, FI-20014, Finland.}

\author{Patrick Rinke}
\email{patrick.rinke@tum.de}

\affiliation[Technical University of Munich]{Department of Physics, Technical University of Munich,
James-Franck-Strasse 1, 85748 Garching,  Germany.}
\alsoaffiliation[Aalto University]
{Department of Applied Physics, Aalto University, P.O. Box 11000, AALTO, FI-00076, Finland.}
\alsoaffiliation[MCML]
{Munich Center for Machine Learning, Germany.}
\alsoaffiliation[AMC]
{Atomistic Modeling Center, Munich Data Science Institute, Technical University of Munich, Walther-von-Dyck Str. 10, 85748 Garching, Germany}

\title{Machine-Learning-Enhanced Mapping of Activity and Selectivity in \ce{CO2} Hydrogenation across Metal-Alloy Surfaces}

\abbreviations{MLFF, OCP, AED, DFT, PCA, VASP, RPBE, EMAE, RWGS}

\begin{document}

\begin{tocentry}

\includegraphics[]{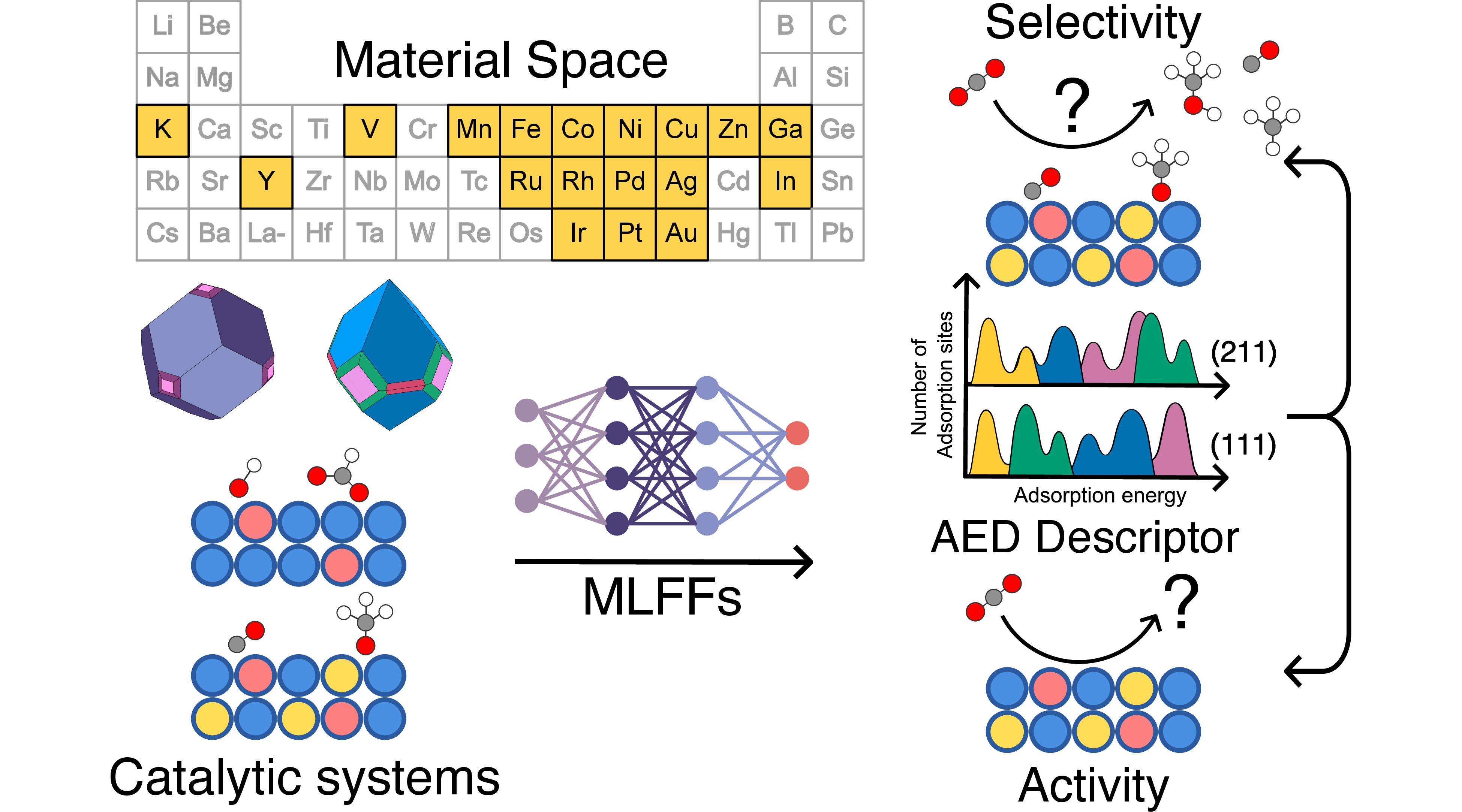}
% Some journals require a graphical entry for the Table of Contents.
% This should be laid out ``print ready'' so that the sizing of the

% Inside the \texttt{tocentry} environment, the font used is Helvetica
% 8\,pt, as required by \emph{Journal of the American Chemical

% The surrounding frame is 9\,cm by 3.5\,cm, which is the maximum
% permitted for  \emph{Journal of the American Chemical Society}
% graphical table of content entries. The box will not resize if the
% content is too big: instead it will overflow the edge of the box.

% This box and the associated title will always be printed on a
% separate page at the end of the document.

\end{tocentry}

\begin{abstract}

Adsorption energy distributions (AEDs) have emerged as a powerful and increasingly adopted descriptor for catalytic performance in high-entropy alloys and, more recently, in conventional metallic alloy nanocrystal catalysts. 
By accounting for diverse adsorption sites and crystallographic facets, AEDs more fully represent nanoparticle-based catalytic surfaces and show strong promise for accelerating rational design and discovery of heterogeneous catalysts, especially for \ce{CO2} hydrogenation.
However, previous high-throughput screenings have not provided simultaneous facet-level insight into catalytic activity, product selectivity, and thermodynamic stability.
Here, we implement and extend the AED framework to individual crystallographic facets, combining facet-resolved adsorption fingerprints with Wulff-derived facet abundances and an interpretable latent-space analysis of C1-product selectivity.
We employ universal machine-learned force fields trained on Open Catalyst data to compute adsorption energies across 226 experimentally observed metals, binary alloys, and previously unexplored ternary alloys, encompassing $\approx$1.4 million adsorption configurations on $>$2,600 crystallographically distinct surfaces. 
Using distribution-based similarity analysis and principal component analysis of AED moments, we identify composition-facet combinations with adsorption landscapes associated with activity and qualitative selectivity trends toward C1 products such as methanol and methane.
Our framework links surface structure to adsorption-based catalytic trends, providing experimentally testable composition-facet candidates for further investigation and validation.

\end{abstract}

\section{Introduction}

Sustainable conversion of sequestered carbon dioxide into methanol over heterogeneous catalysts is a key route toward carbon-neutral chemicals and demands predictive, scalable catalyst design strategies.
Three principal catalytic pathways are pursued for \ce{CO2} hydrogenation: thermal catalysis, electrocatalysis, and photocatalysis~\cite{hu_thermal_2013, mishra_harnessing_2026}. 
Among these, thermal catalysis is currently closest to industrial adaptation due to its technological maturity and compatibility with established large-scale hydrogenation processes~\cite{ye_CO2_2019}.
The development of \ce{CO2}-to-methanol catalysts has been shaped by related industrial processes such as syngas conversion.~\cite{zhang_revealing_2025, cho_advanced_2023, khalil_recent_2025}. 
The industrial standard catalyst for methanol synthesis from both \ce{CO2} and syngas is the multicomponent Cu/ZnO/\ce{Al2O3} system, consisting of metallic and oxide nanoparticles dispersed on an oxide support~\cite{kidd_methanol_1969, hausberger_catalyst_1981}.
These catalysts are often modified with promoters, and their performance is governed by the synergistic contribution of metallic, oxide, and interfacial sites~\cite{chinchen_mechanism_1987, lunkenbein_bridging_2016}.
However, achieving targeted \ce{CO2} conversion remains challenging, since the catalyst design has to simultaneously optimize activity, stability, and product selectivity~\cite{kunkes_hydrogenation_2015, zhou_supercritical_2025, wang_controlling_2023}.

Thermal \ce{CO2} hydrogenation can, in principle, yield a range of valuable C1 products, including formic acid (HCOOH), methane (\ce{CH4}), and methanol (\ce{CH3OH})~\cite{zhang_how_2019, beck_enigma_2024, ye_hydrogenation_2025}.
Methanol, in particular, has attracted attention due to its high energy density, versatility as a chemical feedstock, and compatibility with existing industrial infrastructure~\cite{zhang_how_2019, beck_enigma_2024}.
While methane, formed by the complete hydrogenation of \ce{CO2}, is a gas under standard conditions, it is also often used due to its compatibility with natural gas infrastructure and easier handling than hydrogen~\cite{gotz_renewable_2016, ronsch_review_2016}.
On numerous catalysts, especially Ni-, Co-, and Ru-based, selectivity is governed by competition between methanation and methanol formation~\cite{musab_ahmed_ni-based_2024, godoy-gutierrez_co2_2025, guo_ru_2018}.

Active sites in operating \ce{CO2} hydrogenation catalysts can arise from metallic and alloy nanoparticles, oxide phases, and metal-oxide interfaces, with their identity and reactivity depending on the catalyst composition and reaction environment~\cite{ding_distinctly_2025}.
Ultimately, at the nanoscale, the product formation and selectivity are influenced by the surface structure and electronic properties of the active sites~\cite{zada_disentangling_2026}.
Both the exposed facet and the alloy composition can change the activity and the preferred product~\cite{sterk_structure_2022, brix_tuning_2020, studt_discovery_2014, zimmerli_structure_2024, zhou_reactivity_2024}, so that both have to be resolved to assess the product selectivity of a catalyst.
Experimental screening for such a wide diversity of surface facets with varying compositions is challenging and expensive, thereby requiring scalable computational approaches to catalyst discovery~\cite{hernandez_from_2024, ortega_experimental_2021, bahri_meta-analysis_2022}. 

Theoretical and computational approaches provide a cost-effective route to guide experimental catalyst screening.~\cite{dongapure_mechanistic_2023, gao_recent_2024}
Direct evaluation of catalytic activity and selectivity via reaction barrier calculations remains computationally expensive even for selected adsorption sites and reaction intermediates~\cite{sun_modeling_2025, kreitz_quantifying_2021, polynski_construction_nodate}. 
As a result, many studies rely on simplified descriptors that serve as proxies for catalytic activity. 
Most commonly, these include adsorption energies obtained from density functional theory (DFT)~\cite{medford_sabatier_2015} or more reduced electronic descriptors such as the \textit{d}-band center~\cite{norskov_density_2011, jones_using_2008}.
However, single-site descriptors and the scaling relations between them neglect the site- and facet-heterogeneity inherent to nanoparticle catalysts, limiting their reliability for activity and selectivity trends across complex alloy surfaces~\cite{feng_descriptors_2025, montemore_scaling_2014, andersen_beyond_2019, tan_ai_2025}.
Rational catalyst design, therefore, requires representations that capture the physicochemical complexity of realistic,  nanostructured catalysts.
Recent advances in data-driven approaches~\cite{ouyang_sisso_2018} and machine-learned force fields (MLFFs)~\cite{chanussot_open_2021, kang_large-scale_2020, tran_open_2023, kim_accelerated_2025,tan_ai_2025} offer the computational efficiency needed to sample large adsorption configuration spaces and construct high-fidelity descriptors for heterogeneous catalytic surfaces. 

In our previous work, we leveraged MLFFs~\cite{liao_equiformerv2_2023, tran_open_2023} to construct adsorption energy distributions (AEDs) as a holistic reaction-specific representation of a metallic nanoparticle catalyst. 
By comparing AEDs across a broad space of metals and binary alloys, we identified novel materials with AEDs similar to those of catalysts known to perform well for \ce{CO2}-to-methanol conversion~\cite{pisal_machine_2025}.
Advancing beyond single-site properties, AEDs have provided a compact statistical fingerprint for quantitative comparison of adsorption energies and associated catalytic trends across diverse materials. 

While AEDs naturally account for multiple adsorption sites and, for nanocrystals, multiple facets, they are currently formulated as material-level descriptors.
Moreover, aggregating adsorption energies across all sampled sites implicitly assigns equal weight to each site, irrespective of facet abundance or intrinsic activity, even though catalytic performance is often governed by a limited set of low-index facets~\cite{ye_design_2025, cheng_facet-engineering_2022}. 
In addition, a screening based on the similarity to a single reference catalyst tends to favor materials that resemble the reference, and a single reference does not allow different products, such as methanol and methane, to be distinguished.

Motivated by the demonstrated success of AED-type descriptors in encompassing surface heterogeneity in high-entropy alloy surfaces and explaining their reactivity~\cite{wang_harnessing_2025, pedersen_high-entropy_2020, tan_ai_2025}, we hypothesize that the rich information encoded in AEDs can be leveraged as physically interpretable and statistical descriptors for high-throughput catalyst discovery.
While several data-driven studies have reported correlations between adsorption strengths and product selectivities~\cite{tang_electricity_2020, mok_data-driven_2023, liao_multilayer_2026}, they have examined only a limited range of substrates and adsorption sites, indicating the need for a more thorough exploration. 
Furthermore, to the best of our knowledge, catalytic descriptors that extend beyond a single-site description~\cite{wang_harnessing_2025, pedersen_high-entropy_2020, ouyang_sisso_2018} have not so far been harnessed to assess catalyst activity and product selectivity at the facet level while concurrently enabling high-throughput screening across a wide compositional space of metals, binary alloys, and ternary alloys within one unified framework.

In the present work, we continue to focus on metallic and alloy nanoparticles, which provide a well-defined model space for developing and validating facet-resolved AED descriptors.
However, we aim to adapt our MLFF-driven screening framework~\cite{pisal_machine_2025} to a broader compositional space that extends beyond simple metals and binary systems to more complex and technologically relevant ternary alloys. 
To disentangle the role of surface orientation in catalytic behavior, we construct and analyze facet-resolved AEDs for individual material-facet combinations rather than aggregated material-level descriptors.
We compare the facet-resolved AEDs through the first Wasserstein distance, with reference facets of catalysts known for methanol and methane formation.
In parallel, we evaluate Wulff-derived facet abundances~\cite{ringe_wulff_2011} to estimate the likelihood of facet exposure in realistic nanoparticles. 
By separating catalytic fingerprints from facet exposure probabilities, we can distinguish surfaces that exhibit favorable adsorption characteristics from those that are thermodynamically likely to be present under equilibrium conditions.

Finally, to discern previously uncovered selectivity, we combine facet-resolved AEDs with statistical learning and distribution-based similarity analysis.
Since C1 selectivity is typically pathway-dependent with various possible mechanisms~\cite{cao_selective_2025, zhou_regulating_2026}, we extend our AED space to account for the adsorption energies of \ce{CO} in the analysis. 
Such a framework enables a statistically grounded assessment of catalytic performance across surface facets and provides a scalable route toward extending data-driven catalyst discovery to increasingly complex material spaces.
Beyond identifying promising composition-facet combinations for methanol and methane formation and their exposure likelihood from the presented material space, our approach also aims to establish a practical, scalable and interpretable methodology to relate the adsorption energies and catalytic behavior.

%%%%%%%%%%%%%%%%%%%%%%%%%%%%%%%%%%%%%%%%%%%%%%%%%%%%%%%%%%%%%%%%%%%%%%%%%%%%%%%%%%%%%%%%%%%%
\section{Computational Methods}
\begin{figure}[t!]
    \centering
    \includegraphics[width=0.5\linewidth]{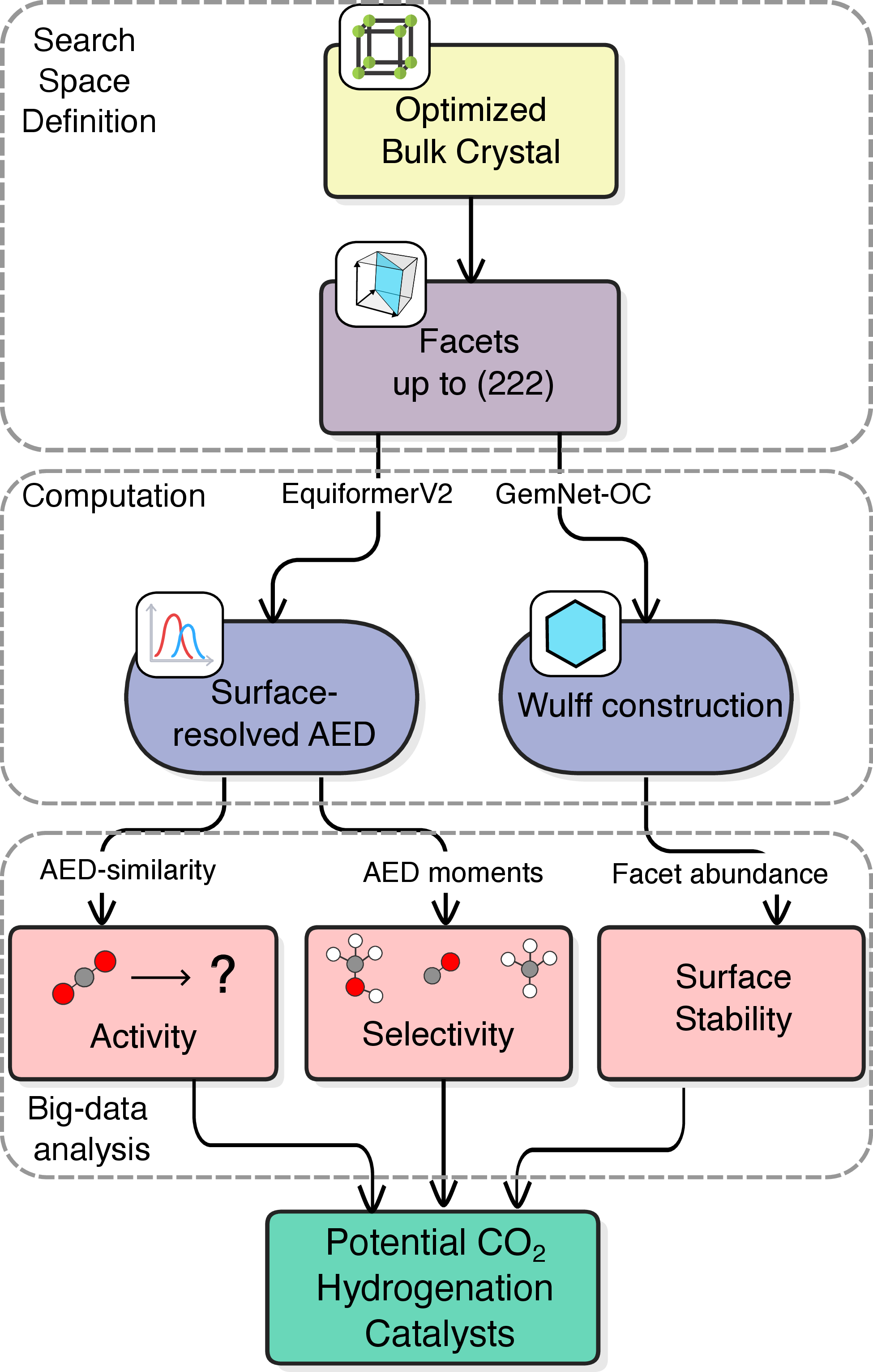}
    \caption{  
    \textbf{Schematic overview of the ML-accelerated facet-wise catalyst screening workflow.} 
    The workflow consists of bulk structure optimization, surface facet generation and stability filtering, preparation of adsorbate-surface configurations, MLFF-driven relaxation to local minima, adsorption energy evaluation, and subsequent data-driven analysis to gain insight into catalytic performance. 
    }
    \label{fig:workflow}
\end{figure}

%%%%%%%%%%%%%%%%%%%%%%%%%%%%%%%%%%%%%%%%%%%%%%%%%%%%%%%%%%%%%%%%%%%%%%%%%%%%%%%%%%%%%%%%%%%%
We adopted and extended the MLFF-accelerated catalyst screening framework developed in our previous work~\cite{pisal_machine_2025}. 
Below, we outline the core workflow for constructing AEDs and elucidate the methodological extensions introduced in this study, including facet-resolved AED analysis, the detection of anomalous adsorbate geometries, Wulff-based facet abundance estimation, and statistical analysis of AEDs.
An overview of the complete workflow is shown in Figure \ref{fig:workflow}.

\subsection{Defining Material Space and Preparing Bulk}

We studied experimentally relevant metallic catalysts composed of transition metals and main-group metals. Our material space comprises the same set of elements as in our previous study, namely the transition metals \{V, Mn, Fe, Co, Ni, Cu, Zn, Y, Ru, Rh, Pd, Ag, Ir, Pt, and Au\} and the main-group metals \{K, Ga, and In\}, and additionally includes Zr in ternary alloys.
Stable and experimentally observed crystal phases for pure metals, binary, and ternary alloys were retrieved from the Materials Project database~\cite{jain_matproj_2013}. Please note that we do not assume a particular crystal phase for any system, as we use the experimentally reported bulk phase.

We optimized bulk geometries using the RPBE functional as implemented in the Vienna Ab Initio Simulation Package (VASP)~\cite{hammer_improved_1999,vasp1_PhysRevB.49.14251,vasp2_PhysRevB.54.11169}. 
During bulk relaxation, the crystallographic symmetry of the reported phase was preserved, and surface models were generated from the corresponding optimized bulk.
We set the plane-wave cutoff to 500 eV for initial relaxations and increased it to 550 eV if necessary for convergence. 
We sampled the Brillouin zone with a k-point spacing of 0.17 $\AA^{-1}$, and automated all calculations using the \texttt{atomate} workflows~\cite{mathew_atomate_2017, jain_fireworks_2015, ong_pymatgen_2013}.

\subsection{Obtaining Stable Surfaces and Wulff Construction}

For each bulk material, we generated all symmetrically distinct crystallographic surfaces with Miller indices -2 $\leq h,k,l \leq$ 2 using the OCP slab-generation workflow implemented in the \texttt{fairchem} toolset~\cite{chanussot_open_2021, fairchem}. For a given triple $(hkl)$, different absolute positions of the Miller plane produce different slab terminations. 
To narrow down the computational space, we evaluated surface energies using thicker slabs of 50 $\AA$ thickness relaxed with the total energy MLFF \texttt{gemnet-oc}~\cite{gasteiger_gemnet-oc_2022}. 
For each Miller index, we retained only the lowest-energy termination for subsequent AED calculations.

Further, we obtained the surface energy using the following equation:

\begin{equation}\label{eq:surface_energy}
    \gamma = \frac{E^{50}_{\text{slab}} - N E_{\text{bulk}}}{2A}
\end{equation}
where $E^{50}_{\text{slab}}$ is the total MLFF-computed energy of the 50 $\AA$ slab, $N$ is the number of slab atoms, $E_{\text{bulk}}$ the bulk energy per atom from DFT relaxation, and $A$ the surface area. For facets that break \textit{z}-directional inversion symmetry, we relaxed both slab terminations, and the average surface energy was considered for the Wulff construction.

We determined the facet proportions in the equilibrium single crystal nanoparticle using the Wulff construction~\cite{ringe_wulff_2011}, implemented in the \texttt{wulffpack}~\cite{rahm_wulffpack_2020} package, based on our surface energy estimates.
When the Wulff construction failed due to slight deviations from ideal crystal symmetry in the relaxed bulk structure, we loosened the symmetry tolerance in \texttt{wulffpack}.
Facets absent from the equilibrium Wulff shape were assigned a Wulff percentage of zero, while facets for materials without a Wulff construction were retained in the screening with an unknown Wulff percentage.

\subsection{Computing Facet-Resolved AEDs}\label{section:relaxtion}

We constructed the production slabs with a thickness of 7~\AA{} and a vacuum spacing of 20~\AA{} along the surface normal to align with the \texttt{fairchem} workflow and the slab construction of the OC20 dataset~\cite{chanussot_open_2021}, on which the MLFFs are trained.
A detailed discussion of the slab thickness is provided in Section~\SIref{sec:slab_tests} of the Supplementary Information (SI), including the MLFF and DFT comparison of 7 and 14~\AA{} slabs (Figure~\SIref{fig:7A_vs_14A_mlff} and Table~\SIref{tab:7A_vs_14A_dft}).

We consider *H (hydrogen atom), *CO (carbon monoxide), *OH (hydroxyl), *OCHO (formate), and *\ce{OCH3} (methoxy) as key surface intermediates in \ce{CO2} hydrogenation, based on previous studies~\cite{tang_CO2_2025, amann_state_2022, zimmerli_how_2025}.
Because the adsorption energies of these intermediates do not strictly follow scaling relations~\cite{montemore_scaling_2014}, we computed the AEDs of each adsorbate explicitly.
Compared with our previous study~\cite{pisal_machine_2025}, the set of adsorbates was extended by *CO.
We sampled high-symmetry adsorption sites and placed the mentioned adsorbates on the surface facets using \texttt{fairchem} tools~\cite{fairchem}, then relaxed each configuration to obtain adsorption energies using the \texttt{equiformerV2} MLFF~\cite{liao_equiformerv2_2023}. We calculated adsorption energies ($E_{ads}^{DFT}$ ) according to:

\begin{equation}
    E_{ads}^{DFT} = E_{system}^{DFT} - E_{surface}^{DFT} - E_{adsorbate}^{DFT}
\end{equation}

Following the convention of Chanussot et al.~\cite{chanussot_open_2021}, we computed spin-unpolarized atomic reference energies using VASP and using linear combinations we obtained the gas phase energies of the adsorbates ($E_{adsorbate}^{DFT}$): *H: --3.489 eV, *OH: --10.690 eV, *CO: --14.483 eV, *\ce{OCH3}: --24.950 eV, *OCHO: --25.173 eV.

We identified unique optimized geometries for each slab termination to avoid over-representation of identical local minima.
A configuration was considered a duplicate if all adsorbate atoms lay within 0.1~\AA{} of the corresponding atoms in any previously accepted configuration on the same slab termination.
In contrast to material-specific AEDs in the previous workflow~\cite{pisal_machine_2025}, we obtained surface-resolved AEDs by aggregating the unique adsorption energies per Miller surface and adsorbate into a probability distribution on a 0.1~eV grid from --15 to 5~eV.
For materials with asymmetric terminations, we aggregated adsorption energies to obtain a unified AED that represents both terminations for each Miller index.

In comparison with the previous workflow~\cite{pisal_machine_2025}, we additionally checked each relaxed configuration for anomalous adsorbate geometries, since *CO frequently desorbed from the surface during the MLFF relaxations and several *OCHO configurations dissociated.
A configuration was classified as desorbed if no adsorbate--surface atom pair lay within 1.5 times the sum of the covalent radii of the two atoms.
Within the adsorbate, two atoms were considered bonded if their distance was below 1.6~\AA{} for C--H and below 2.0~\AA{} for C--O and for O--H in *OH.
A configuration was classified as dissociated if the resulting bond graph split into disconnected fragments.
We retained the anomalous configurations in the AEDs as separate bins above the range of regular adsorption energies: desorption at 5.0~eV, C--H cleavage at 5.5~eV, C--O cleavage at 6.0~eV, and O--H cleavage at 6.5~eV.
The fraction of configurations that do not remain adsorbed or intact therefore contributes to the distance between AEDs.
This leads to a final AED range of --15 to 7~eV, and the AEDs are normalized to one for each adsorbate.

To validate the MLFF predictions, we selected three configurations for each material-adsorbate pair corresponding to the minimum, median, and maximum adsorption energy and performed spin-unpolarized single-point DFT calculations on these structures and the corresponding MLFF-optimized facets with a plane-wave cutoff of 450 eV.
The estimated mean absolute error (EMAE) of the MLFF predictions for each material-adsorbate pair was computed as:

\begin{equation}
    \mathrm{EMAE} = \frac{1}{n} \sum_{i=1}^{n} \left| E_{\mathrm{ads},i}^{\mathrm{DFT}} - E_{\mathrm{ads},i}^{\mathrm{eq2}} \right|,
\end{equation}
where $n=3$ corresponds to the three selected configurations (minimum, median, maximum) and $E_{\mathrm{ads},i}^{\mathrm{eq2}}$ denotes the adsorption energy predicted by the MLFF for the $i^{th}$ configuration. Similarly to our previous work~\cite{pisal_machine_2025}, we averaged the EMAE over the five adsorbates of each material and discarded a material from the analysis if this average exceeded 0.25~eV.

\subsection{Modeling the Reference Catalysts}

\begin{figure}
    \centering
    \includegraphics[width=\linewidth]{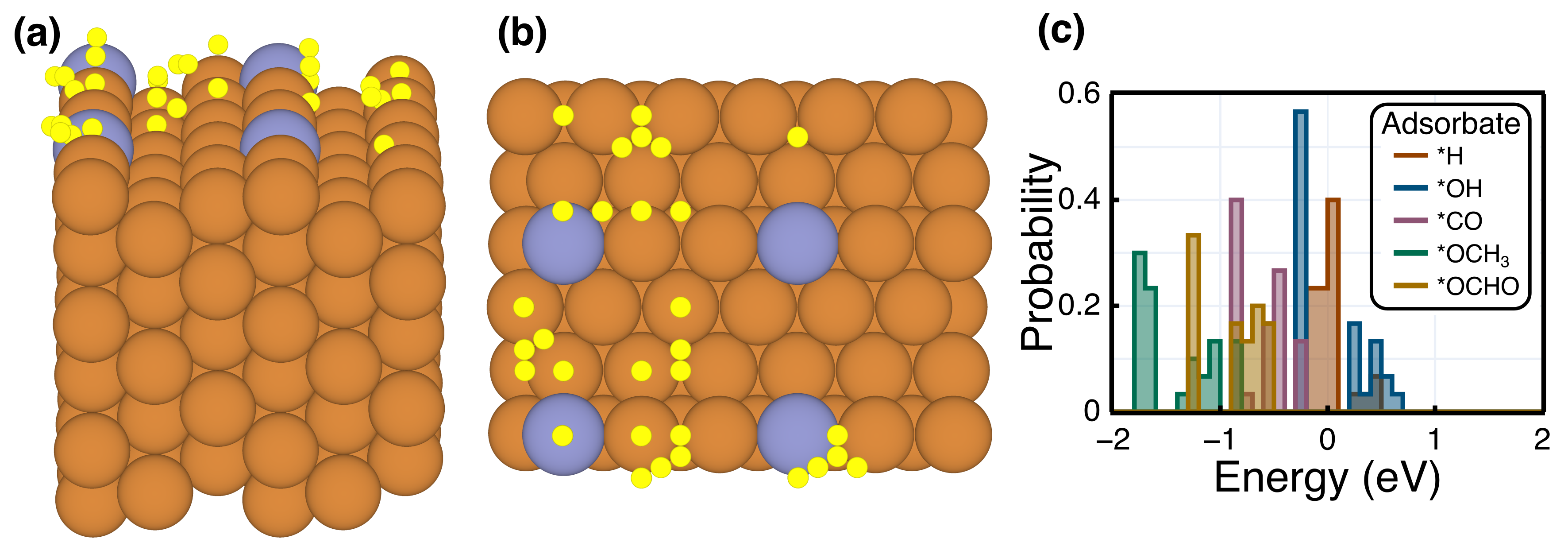}
    \caption{(a) Side view and (b) top view of the Zn@Cu(211) reference facet. Orange spheres indicate Cu atoms, purple spheres represent Zn atoms, and yellow points denote high-symmetry adsorption sites identified by the sampling algorithm. (c) AEDs for the Zn@Cu(211) facet, capturing the adsorption energies over the highlighted adsorption sites.
    }
    \label{fig:Zn@Cu(211)}
\end{figure}

We selected a Cu-Zn catalyst as the reference system for methanol synthesis due to its well-characterized experimental performance for \ce{CO2} hydrogenation to methanol by
Behrens et al. (2012)~\cite{behrens_active_2012} and Amann et al. (2022)~\cite{amann_state_2022}, with both studies providing detailed insight into Cu(211) surfaces with substituted Zn atoms.
The promotion of Cu step sites by Zn has further been quantified experimentally and described mechanistically~\cite{kuld_quantifying_2016, studt_mechanism_2015}.
To simulate the Cu-Zn system in the same framework as the rest of our workflow, we built the Cu(211) system using optimized Cu(fcc) bulk and the \texttt{fairchem} workflow. 
To ensure sufficient separation between periodic images of the adsorbates, we constructed a 6$\times$2 supercell of the (211) slab, in which every third atom along the upper step edge of the surface was substituted with Zn atoms.
The procedure yielded a total of four Zn atoms on the 192-atom Cu(211) slab as shown in Figure~\ref{fig:Zn@Cu(211)}(a) and (b). 
We refer to this system as Zn@Cu(211). 
High-symmetry adsorption sites identified via the \texttt{fairchem} workflow are highlighted in Figure~\ref{fig:Zn@Cu(211)}. 
The AED descriptor for this particular system, computed using \texttt{equiformerV2}, is shown in Figure~\ref{fig:Zn@Cu(211)}(c).

As the reference system for \ce{CO2} methanation, we selected the Ni(110) facet.
Based on DFT calculations and microkinetic modeling, Sterk et al.~\cite{sterk_structure_2022} identified Ni(110) as the most active Ni facet for \ce{CO2} methanation, whereas the Ni(111), Ni(100), and Ni(211) facets showed little activity.
Near-ambient-pressure surface studies likewise identify Ni(110) as the most active Ni facet for \ce{CO2} methanation~\cite{roiaz_reverse_2016, zang_co2_2023}.
Since Ni(110) is part of the screened material space, we used the Ni(110) AEDs computed in the same workflow as all other facets.

\subsection{AED Similarity and Latent Space Representation}

We quantified the similarity between facet-resolved AEDs with the first Wasserstein distance ($W_1$)~\cite{virtanen_scipy_2020, ramdas_wasserstein_2017}.
For each adsorbate, we computed the $W_1$ between the binned AEDs of two facets, using the bin centers as the support and the AED probabilities as the weights.
We defined the distance $W_1$ between two facets as the mean of the per-adsorbate distances over the five adsorbates.
Throughout this work, $W_{\mathrm{Cu}}$ and $W_{\mathrm{Ni}}$ denote the distance of a facet to the Zn@Cu(211) and Ni(110) references, respectively.
We analyzed the facets around each reference at two distance thresholds.
The inner threshold of $W_1 \leq 0.25$~eV equals the EMAE threshold used to accept the MLFF predictions, so that the differences between the facets within this threshold and the reference remain comparable to the estimated MLFF error.
To extend the search toward facets with a less similar but related adsorption landscape, we additionally considered all facets with $W_1 \leq 0.40$~eV to a reference.

To further probe the differences between AEDs, we used two additional methods:
Uniform Manifold Approximation and Projection for Dimension Reduction (UMAP)~\cite{mcinnes_umap_2020} and
Principal Component Analysis (PCA)~\cite{mackiewicz_principal_1993}.
UMAP can capture nonlinear relationships and preserve local neighborhood structure in its 2D projection, but its axes and embedding directions do not provide direct physical interpretation.
We computed the UMAP embedding from the concatenated AED probabilities of the five adsorbates on the regular energy grid (1,000 features per facet) with the Euclidean metric, 15 nearest neighbors, and a minimum distance of 0.1.

PCA projection to the lower-dimensional latent space using a simplified representation of the AEDs provides further insight.
For that we computed statistical moments of the unique regular adsorption energies for each facet-adsorbate pair, including the mean, median, standard deviation (std), 5\textsuperscript{th}, 25\textsuperscript{th}, 75\textsuperscript{th}, and 95\textsuperscript{th} percentiles, minimum (min), maximum (max), skewness, and kurtosis.
We performed the PCA without standardizing the moments, so that all moments retain their common energy scale in eV.
Facets on which all *CO configurations were anomalous have no regular *CO adsorption energies and therefore no *CO moments, and these facets were excluded from the PCA.
For the final PCA, we used the minimum, 5\textsuperscript{th} percentile, mean, median, maximum, and std of each adsorbate (30 features).
We chose this set of moments to maximize the variance explained by the first two principal components while keeping the PCA embedding similar to the UMAP embedding.
We used these PCA loading vectors to evaluate interpretable trends in the AED distributions and relate them to the catalytic behavior of the screened surfaces.

%%%%%%%%%%%%%%%%%%%%%%%%%%%%%%%%%%%%%%%%%%%%%%%%%%%%%%%%%%%%%%%%%%%%%%%%%%%%%%%%
\section{Results}
%%%%%%%%%%%%%%%%%%%%%%%%%%%%%%%%%%%%%%%%%%%%%%%%%%%%%%%%%%%%%%%%%%%%%%%%%%%%%%%%

\subsection{Dataset and Surface-Resolved AEDs}

We computed 226 stable experimentally observed materials retrieved from the Materials Project Database, comprising 16 pure metals (7\%), 149 binary alloys (66\%), and 61 ternary alloys (27\%). 
The complete list of bulk systems is provided in Table~\SIref{tab:wmat} of the SI. 
These materials yielded 4,841 unique slab terminations, which correspond to 2,613 crystallographically distinct Miller planes, of which 132 (5\%), 1,705 (65\%), and 776 (30\%) belong to pure metals, binary alloys, and ternary alloys, respectively. 
Placing the five adsorbates on the surface slabs via \texttt{fairchem} yielded approximately 1.4 million surface-adsorbate configurations, which we relaxed using \texttt{equiformerV2} trained on the OC20 dataset with reported accuracy of 0.23 eV for adsorption energy predictions~\cite{liao_equiformerv2_2023}. 

We validated MLFF-predicted adsorption energies through the min-max-median sampling technique described above. 
Across all material-adsorbate pairs, the resulting EMAE was 0.11 eV, which falls below the reported accuracy of the pretrained model. 
The prediction plot and the distribution of EMAE across the material space are shown in the SI (Figure~\SIref{fig_predictions}).
The *H adsorption energies on 7 and 14~\AA{} slabs of the representative VPt alloy differed by at most 0.04~eV in DFT, which confirms that the slab thickness of 7~\AA{} does not limit the accuracy of the AEDs (Section~\SIref{sec:slab_tests}, Figure~\SIref{fig:7A_vs_14A_mlff}, and Table~\SIref{tab:7A_vs_14A_dft} of the SI).

The check for anomalous adsorbate geometries identified 20,286 desorbed *CO configurations, which correspond to 7.3\% of all *CO configurations and occur on facets of 148 of the 226 materials.
Dissociation was less frequent and mainly affected *OCHO, with 3,924 configurations showing C--H cleavage, whereas all other anomaly categories together contained fewer than 300 configurations.
The Zn@Cu(211) and Ni(110) references contained no anomalous configurations, and the highest regular adsorption energy across all facets was 3.69~eV, below the anomaly bins at 5.0 to 6.5~eV.
The classification of the anomalous configurations and example AEDs with anomaly bins are provided in Section~\SIref{sec:anomalies} and Figure~\SIref{fig:anomaly_bins} of the SI.

The aggregated AED for Zn@Cu(211) is shown in Figure \ref{fig:Zn@Cu(211)}(c), and 
several examples of AEDs are shown in Figures~\SIref{fig:anomaly_bins}--\SIref{fig_closest_Ni110} of the SI.
In contrast to the broad, continuous distributions observed in material-aggregated AEDs~\cite{pisal_machine_2025}, these facet-resolved profiles are characterized by discrete, discontinuous peaks.
Figure~\SIref{fig:aed_groups} of the SI shows the group-mean AEDs, i.e., the AEDs averaged over all facets of the materials in each chemistry group (e.g., materials containing Fe, Co, or Ni), with each facet weighted equally.
Among the considered intermediates, *\ce{OCH3} has the lowest median adsorption energy on 86\% of the facets.
In contrast, *H and *OH have a positive median adsorption energy on 32\% and 24\% of the facets, respectively, and their relative position is highly material and facet-dependent.
Please note that positive values do not necessarily imply desorption, as our calculations neglect entropy, zero-point corrections, or pressure contributions.
Overall, the relative ordering and spread of adsorption energies across adsorbates vary strongly between facets, indicating that simple universal trends are not easily established from single-value descriptors.

Notably, *CO adsorption energies exhibit a strong dependence on the material, with median values ranging from --3.38 to 0.28~eV across the facets, and *CO has the lowest median adsorption energy on 12\% of the facets.
The AEDs of all adsorbates depend on the chemistry of the material (Figure~\SIref{fig:aed_groups} of the SI).
For *CO, the median adsorption energy of the group-mean AEDs shifts from about --1.6~eV for materials containing Fe, Co, Ni, Ru, Rh, or Ir, through --1.1~eV for materials containing Pd or Pt, to --0.6~eV for Cu-, Ag-, and Au-based materials and --0.3~eV for materials containing only Zn, Ga, In, or K.
Among the main-group metals, 41\% of *CO configurations desorb.
The ordering of the groups differs between the adsorbates: materials containing early transition metals (Y, V, Zr, Mn) bind *CO only moderately (--0.9~eV), but bind *OH and *\ce{OCH3} most strongly (--1.2 and --2.9~eV) and *OCHO as strongly as the materials containing Ru, Rh, or Ir (--2.0 and --2.1~eV).

\subsection{Facet-wise Comparison and Facet Stability}

\begin{figure}
    \centering
    \includegraphics[width=\linewidth]{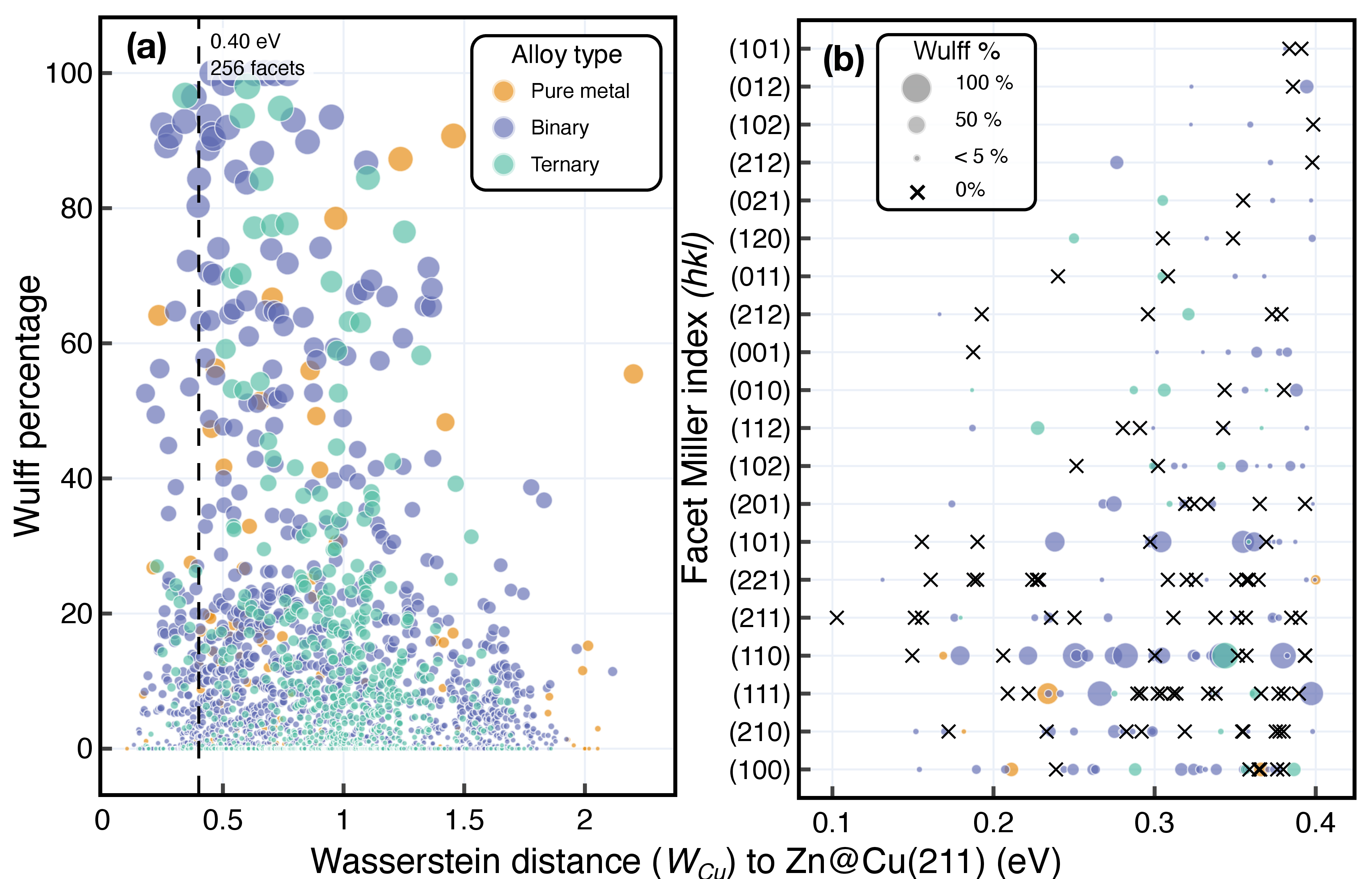}
    \caption{(a) Scatter plot of the Wasserstein distance $W_{\mathrm{Cu}}$ to the Zn@Cu(211) reference facet versus Wulff percentage for all 2,613 catalyst facets. The facets are colored by system type, and the marker size reflects the Wulff percentage. The vertical dashed line at $W_{\mathrm{Cu}} = 0.40$~eV marks the threshold of the extended search, which includes 256 facets. (b) $W_{\mathrm{Cu}}$ of the facets within the extended threshold, grouped by Miller index, with marker size denoting the Wulff percentage. For clarity, only Miller indices with at least three facets within the extended threshold are shown (20 of 30 Miller indices, 239 of 256 facets). Cross markers ($\times$) represent facets that are not part of the equilibrium Wulff shape in vacuum.}
    \label{fig:wsd_vs_facets}
\end{figure}

To compare all catalyst facets, we computed $W_{\mathrm{Cu}}$ between each facet and the Zn@Cu(211) reference.
Figure~\ref{fig:wsd_vs_facets}(a) presents a scatter plot of $W_{\mathrm{Cu}}$ versus the Wulff percentage for all 2,613 distinct facets, with $W_{\mathrm{Cu}}$ ranging from 0.103 to 2.20~eV.
Each facet is colored according to its system type: pure metal, binary alloy, or ternary alloy. 

The size of the marker indicates the Wulff percentage.
Of the 2,518 facets with a Wulff construction, 804 facets (32\%) do not appear on the equilibrium Wulff shape, and the number of facets decreases considerably with increasing abundance.
Among the 1,714 facets that appear on the equilibrium Wulff shape, the Wulff percentage is not correlated with $W_{\mathrm{Cu}}$ or $W_{\mathrm{Ni}}$ (Spearman $\rho = -0.07$ and $-0.02$, respectively).
The fraction of facets within the extended threshold ($W_1 \leq 0.40$ eV) is similar for facets that do not appear and facets that appear on the Wulff shape (11.7\% and 9.5\% for Zn@Cu(211), 5.1\% and 5.8\% for Ni(110)).
For six materials, the Wulff construction was obtained only after loosening the symmetry tolerance.
For nine materials with a total of 95 facets, no Wulff construction could be obtained owing to non-positive surface energies for six materials and \texttt{wulffpack} failures for three materials.

Within the inner threshold of $W_{\mathrm{Cu}} \leq 0.25$~eV, we find 53 facets of 13 materials.
The closest facet is Cu(211) with $W_{\mathrm{Cu}} = 0.103$~eV, followed by \ce{CuZn}(221), \ce{GaCu3}(110), \ce{Cu3Au}(211), and \ce{CuZn}(210), indicating that the distance measure recovers the Cu-based chemistry of the reference.
With the exception of \ce{K(InAu2)2}(101) and GaPd(210), all 53 facets within the inner threshold contain either Cu or Zn.
To look beyond Cu- and Zn-based materials, we extended the search to $W_{\mathrm{Cu}} \leq 0.40$~eV, which includes 256 facets of 57 materials (Table~\SIref{tab:selected_facets} of the SI).
Between the two thresholds, 63 of the 203 additional facets contain neither Cu nor Zn, and 44 materials appear that are absent within the inner threshold.
Within the extended threshold, binary alloy facets predominate (201 facets, 79\%), followed by ternary alloys (40 facets, 16\%) and pure metals (15 facets, 6\%).

Figure~\ref{fig:wsd_vs_facets}(b) shows the facet-wise distribution of $W_{\mathrm{Cu}}$ across the facets within the extended threshold, with the Miller indices indicated on the \textit{y}-axis. 
The most commonly observed Miller index is (100) with 29 facets, followed by (210) and (111) with 26 facets each.
Of the 256 facets, 162 facets appear on the equilibrium Wulff shape.
Among the 20 facets closest to the reference, seven facets do not appear on the Wulff shape and four facets have a Wulff percentage below 1\%.
The most abundant facet among these 20 facets is \ce{Zn11Rh2}(110), which accounts for 52.6\% of the Wulff shape, and the most abundant facet within the extended threshold is \ce{Zn2AgAu}(110) with 96.7\% at $W_{\mathrm{Cu}} = 0.344$~eV.
The overall distribution emphasizes low-index facets, with 22 facets featuring negative Miller indices.

We performed the same analysis for the Ni(110) methanation reference, which lies at $W_1 = 0.491$~eV from Zn@Cu(211).
Within the inner threshold of $W_{\mathrm{Ni}} \leq 0.25$~eV, we find 28 facets of 16 materials.
The closest facets are Ni(210) with $W_{\mathrm{Ni}} = 0.127$~eV, \ce{Ni2Zn2}(211) with 0.148~eV, and \ce{Ni3Pt}(110) with 0.154~eV, and 21 of the 28 facets contain Ni.
The extended threshold of $W_{\mathrm{Ni}} \leq 0.40$~eV includes 141 facets of 40 materials (Table~\SIref{tab:selected_facets_Ni110} of the SI), and 54 of the 113 additional facets do not contain Ni.
These facets comprise 100 binary alloy facets (71\%), 28 pure metal facets (20\%), and 13 ternary alloy facets (9\%).
The facets around the two references do not overlap within the inner threshold.
Within the extended threshold, 27 facets lie close to both references, of which 17 facets contain Ni, including eight \ce{Ni2Zn2} and three NiZn facets, together with facets of \ce{Ni3Pt}, \ce{Ga3Ni5}, and Rh.

\begin{table}[htbp]
    \centering
    \caption{Facets closest to the methanol reference Zn@Cu(211), excluding Cu- and Zn-containing materials, and to the methanation reference Ni(110), excluding Ni-containing materials, in ascending order of the distance to the respective reference (shown in bold).
    For each facet, the table lists the Materials Project ID, the Miller index, the Wulff percentage, the coordinates in the 2D PCA latent space, and the distances $W_{\mathrm{Cu}}$ and $W_{\mathrm{Ni}}$ to both references.
    The corresponding AEDs are shown in Figures~\SIref{fig_closest_noCuZn} and~\SIref{fig_closest_Ni110} of the SI.}
    \begin{tabular}{|l l l r r r r r|}
\toprule
\textbf{Material} & \textbf{mp-ID} & \textbf{$(hkl)$} & \textbf{\shortstack{Wulff\\Percentage}} & \textbf{PC1} & \textbf{PC2} & \textbf{$W_{\mathrm{Cu}}$ (eV)} & \textbf{$W_{\mathrm{Ni}}$ (eV)} \\
\midrule
\multicolumn{8}{|l|}{\textbf{Zn@Cu(211) (\ce{CH3OH}), excluding Cu- and Zn-containing materials}} \\
\midrule
\ce{K(InAu2)2} & 21134 & (101) & 0.00 & 4.00 & 1.11 & \textbf{0.190} & 0.638 \\
\ce{GaPd} & 1078526 & (210) & 0.00 & 3.76 & 0.44 & \textbf{0.233} & 0.659 \\
\ce{K(InAu2)2} & 21134 & (111) & 3.58 & 3.90 & 1.39 & \textbf{0.275} & 0.719 \\
\ce{Ni2Ga3} & 11397 & (2$\bar{1}$2) & 25.25 & 2.70 & -0.49 & \textbf{0.277} & 0.452 \\
\ce{YGa2Pd} & 1079075 & (010) & 9.06 & 2.39 & 1.30 & \textbf{0.287} & 0.654 \\
\ce{K(InAu2)2} & 21134 & (112) & 0.00 & 4.39 & 0.87 & \textbf{0.291} & 0.729 \\
\ce{K(InAu2)2} & 21134 & (102) & 6.44 & 4.35 & 1.11 & \textbf{0.299} & 0.766 \\
\ce{KAu5} & 1298 & (102) & 3.61 & 4.65 & 0.43 & \textbf{0.299} & 0.761 \\
\ce{KAu2} & 30401 & (101) & 64.75 & 2.58 & 1.30 & \textbf{0.304} & 0.473 \\
\ce{K3InAu5} & 867331 & (011) & 11.39 & 2.77 & 1.71 & \textbf{0.305} & 0.550 \\
\ce{K3InAu5} & 867331 & (010) & 24.26 & 2.93 & 1.90 & \textbf{0.306} & 0.624 \\
\midrule
\multicolumn{8}{|l|}{\textbf{Ni(110) (\ce{CH4}), excluding Ni-containing materials}} \\
\midrule
\ce{VIr} & 569250 & (010) & 5.16 & 1.35 & -1.33 & 0.475 & \textbf{0.183} \\
\ce{Ru} & 33 & (100) & 19.82 & 2.28 & -1.99 & 0.441 & \textbf{0.191} \\
\ce{Ru} & 33 & (001) & 17.40 & 2.77 & -1.93 & 0.412 & \textbf{0.212} \\
\ce{CoPt} & 949 & (211) & 10.06 & 2.58 & -1.84 & 0.464 & \textbf{0.249} \\
\ce{YPt3} & 2403 & (111) & 63.38 & 2.23 & -1.20 & 0.448 & \textbf{0.249} \\
\ce{FePt} & 2260 & (110) & 6.05 & 3.18 & -2.02 & 0.454 & \textbf{0.265} \\
\ce{FePt} & 2260 & (210) & 0.00 & 1.74 & -1.63 & 0.510 & \textbf{0.270} \\
\ce{Rh} & 74 & (100) & 14.34 & 3.15 & -2.29 & 0.456 & \textbf{0.272} \\
\ce{MnGaCo2} & 21171 & (111) & 3.22 & 2.07 & -0.91 & 0.360 & \textbf{0.279} \\
\ce{Co} & 54 & (211) & 0.01 & 1.36 & -1.92 & 0.633 & \textbf{0.286} \\
\ce{Ru} & 33 & (101) & 41.70 & 2.51 & -2.06 & 0.504 & \textbf{0.287} \\
\bottomrule
\end{tabular}
 
    \label{tab:selectivity_map}
\end{table}

Table~\ref{tab:selectivity_map} lists the facets closest to each reference after excluding the materials that share the main constituents of the reference.
For Zn@Cu(211), the list is dominated by K--Au--In-based facets, with \ce{K(InAu2)2}(101) as the closest facet at $W_{\mathrm{Cu}} = 0.190$~eV, followed by GaPd(210), \ce{K(InAu2)2}(111), \ce{Ni2Ga3}(2$\bar{1}$2), and \ce{YGa2Pd}(010).
Among these facets, \ce{KAu2}(101) and \ce{Ni2Ga3}(2$\bar{1}$2) are abundant on the equilibrium Wulff shape with 64.8\% and 25.3\%, respectively.
For Ni(110), the list contains three Ru facets, with Ru(100) as the second-closest facet at $W_{\mathrm{Ni}} = 0.191$~eV, together with VIr(010), CoPt(211), \ce{YPt3}(111), FePt, Rh(100), \ce{MnGaCo2}(111), and Co(211).
\ce{YPt3}(111) and Ru(101) are abundant on the equilibrium Wulff shape with 63.4\% and 41.7\%, respectively.

\subsection{AED Projection on 2D Latent Space}

\begin{figure}[t!]
    \centering
    \includegraphics[width=\linewidth]{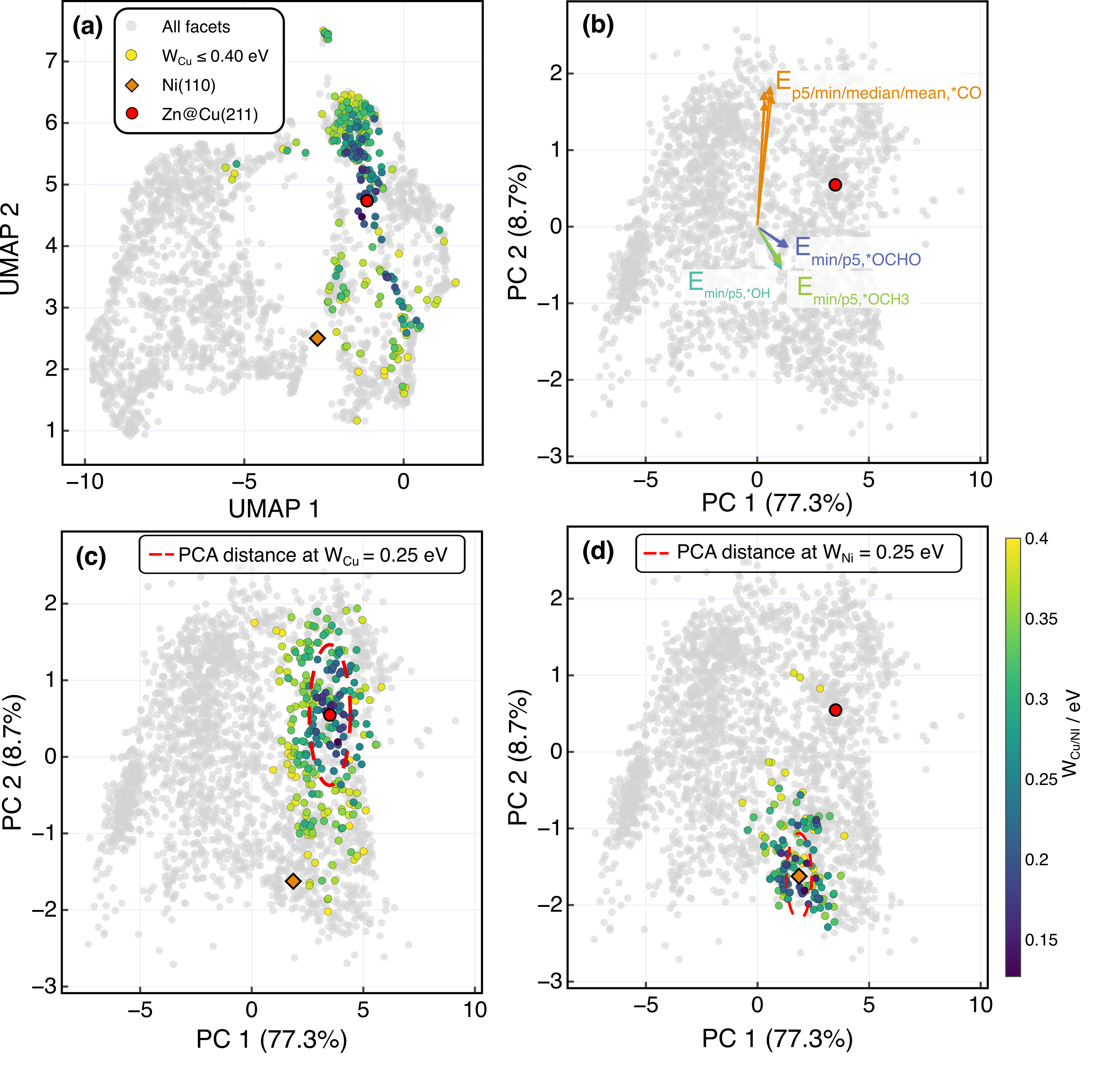}
    \caption{(a) UMAP embedding of the AEDs of all facets. The facets within $W_{\mathrm{Cu}} \leq 0.40$~eV are colored by $W_{\mathrm{Cu}}$, and the Zn@Cu(211) and Ni(110) references are highlighted.
    (b) PCA biplot of the six selected AED moments, showing the directionality and contribution of the dominant moments to each principal component. The gray data points represent the 2,609 facets included in the PCA. The loading vectors have been scaled for visibility and do not represent absolute magnitudes.
    (c) PCA latent space with the facets within $W_{\mathrm{Cu}} \leq 0.40$~eV of Zn@Cu(211), colored by $W_{\mathrm{Cu}}$.
    (d) PCA latent space with the facets within $W_{\mathrm{Ni}} \leq 0.40$~eV of Ni(110), colored by $W_{\mathrm{Ni}}$.
    In (c) and (d), the dashed circle indicates the PCA distance of the facet at the inner threshold of $W_1 = 0.25$~eV.
    }
    \label{fig:pca}
\end{figure}

To better understand the information encoded in the AEDs and relate the AEDs to the catalytic behavior of the facets, we projected the AEDs onto a lower-dimensional space.
Figure~\ref{fig:pca}(a) shows the UMAP embedding of the AEDs of all facets.
The facets within $W_{\mathrm{Cu}} \leq 0.40$~eV are concentrated in a region of the UMAP embedding around Zn@Cu(211), with a few facets in other parts of the embedding, whereas Ni(110) lies outside the region around Zn@Cu(211).
While the UMAP embedding cannot be interpreted directly, the PCA of the AED moments provides physically interpretable axes.
For the PCA, we excluded four facets without any regular *CO configuration, leaving 2,609 facets in addition to the reference.

PCA on the six selected moments revealed that the first principal component (PC1) accounts for 77.3\% of the explained variance, while the second principal component (PC2) captures 8.7\%, for a total of 86.1\% across the dataset. 
In contrast, a PCA of all eleven moments, including skewness and kurtosis, explains only 31.7\% and 21.7\% of the variance with PC1 and PC2, respectively, and the corresponding loading vectors are dominated by the kurtosis (Figure~\SIref{fig_umap} of the SI).
The facet-resolved AEDs, their statistical moments, the PCA coordinates, and the distances $W_{\mathrm{Cu}}$ and $W_{\mathrm{Ni}}$ of all facets are provided on the Zenodo platform~\cite{pisal_dataset_2026}.

Figure~\ref{fig:pca}(b) illustrates the 2D projection of the dataset, including the top 10 loading vectors. 
The corresponding loading values for PC1 and PC2 are shown in Figure~\SIref{fig:relative_contri_pca} of the SI.
Note that these vectors have been scaled for visibility and do not represent absolute magnitudes. 
PC1 is dominated by the minimum and the 5\textsuperscript{th} percentile of the O-bound intermediates *OCHO, *OH, and *\ce{OCH3}, which adsorb through an O atom, with the largest contributions from the minimum (11.6\%) and the 5\textsuperscript{th} percentile (10.0\%) of *OCHO, followed by the corresponding moments of *OH (6.9\%) and *\ce{OCH3} (6.7\%).
Since the loadings of these moments are positive, larger PC1 values correspond to weaker binding of the O-bound intermediates.
PC2 receives 77\% of its contribution from *CO moments, mainly from the 5\textsuperscript{th} percentile (19.2\%), the median (17.7\%), the minimum (16.9\%), and the mean (15.1\%), so that larger PC2 values correspond to weaker *CO binding.
The standard deviation contributes less than 0.5\% to either principal component, and the maximum contributes mainly through *CO to PC2 (7.3\%).
The two references are separated along both principal components, with Ni(110) at lower PC1 and PC2 values than Zn@Cu(211).

Figures~\ref{fig:pca}(c) and (d) show the facets within the extended threshold of each reference in the PCA latent space.
The facets within the inner threshold remain near the respective reference, with a median PCA distance of 0.64 to Zn@Cu(211) and 0.68 to Ni(110).
As $W_{\mathrm{Cu}}$ increases toward 0.40~eV, the facets scatter in all directions around Zn@Cu(211).
For the facets with $0.35 < W_{\mathrm{Cu}} \leq 0.40$~eV, the median PCA distance increases to 1.69, and the standard deviations of PC1 and PC2 increase from 0.59 to 1.20 and from 0.42 to 1.04, respectively, relative to the facets within the inner threshold.
Around Ni(110), the scatter is less pronounced, with a median PCA distance of 0.97 for the facets with $0.35 < W_{\mathrm{Ni}} \leq 0.40$~eV, and the standard deviations of PC1 and PC2 increase from 0.61 to 1.00 and from 0.31 to 0.75, respectively.
The radial scatter shows that the facets beyond the inner threshold deviate from the reference in different directions in the PCA space.
Although the facets in the extended neighborhood exhibit more varied and novel compositions, the interpretation of the PCA distance from their respective references warrants a more thorough discussion.

%%%%%%%%%%%%%%%%%%%%%%%%%%%%%%%%%%%%%%%%%%%%%%%%%%%%%%%%%%%%%%%%%%%%%%%%%%%%%%%%
\section{Discussion}
%%%%%%%%%%%%%%%%%%%%%%%%%%%%%%%%%%%%%%%%%%%%%%%%%%%%%%%%%%%%%%%%%%%%%%%%%%%%%%%%

\subsection{AED-based Activity and Its Assessment Limitation}

In this study, we performed facet-wise screening of catalytic candidates using AEDs of relevant reaction intermediates as descriptors. 
In contrast to our previous facet-averaged workflow, which aggregated adsorption energies over multiple thermodynamically stable low-index surfaces~\cite{pisal_machine_2025}, we now resolve and analyze each crystallographic facet individually and quantify its apparent catalytic activity based on the similarity of its AED to that of a reference facet.
Specifically, we use the first Wasserstein distance between facet-resolved AEDs as a measure of proximity in adsorption energy space.
Facets with a smaller $W_{\mathrm{Cu}}$ are considered more likely to reproduce the adsorption landscape of Zn@Cu(211) associated with methanol formation, and facets with a smaller $W_{\mathrm{Ni}}$ are considered more likely to reproduce the adsorption landscape of Ni(110) associated with methanation.
Beyond providing a structural fingerprint, AEDs enable the comparison of apparent facet activity across a diverse material space. 
Within the extended threshold, the fingerprint aids us in identifying facets that are compositionally different from the references yet possess similar AEDs (Tables~\SIref{tab:selected_facets} and~\SIref{tab:selected_facets_Ni110} of the SI) and, subsequently, potential activity toward \ce{CO2} hydrogenation.

Our analysis highlights both the utility and limitations of the $W_1$ as a similarity metric for screening catalysts, suggested in our previous study~\cite{pisal_machine_2025}.
$W_1$ is sensitive to both the probability distribution and the positions of the corresponding AED values, and therefore captures differences in distribution shape as well as systematic shifts in adsorption energies. 
However, $W_1$ generalizes the complex AED landscape into a single number and thus cannot fully capture whether the differences are caused by a systematic shift or by a subset of the adsorption sites.

The PCA embeddings in Figures~\ref{fig:pca}(c) and (d) illustrate this limitation, since facets at a similar distance from a reference deviate from the reference in different directions of the latent space.
Conversely, facets close to a reference in the PCA space can be distant in terms of $W_1$.
Of the 72 facets within a PCA distance of 0.5 from Zn@Cu(211), 48 facets have $W_{\mathrm{Cu}} > 0.40$~eV, and on 45 of these 48 facets at least 20\% of the unique *CO configurations desorbed.
In contrast, the median *CO desorption fraction of the remaining 24 facets in the same region is 1\%.
Since the moments exclude the anomalous configurations, whereas $W_1$ accounts for the anomaly bins, the two measures provide complementary information, and $W_1$ penalizes facets on which a large fraction of *CO does not remain adsorbed.
For example, \ce{In3Ni2}(001) belongs to the In--Ni compositions identified as promising for methanol synthesis in previous studies~\cite{khatamirad_data-driven_2023, zhu_ni-synergy_2021} and lies close to Zn@Cu(211) in the PCA space.
However, *CO desorbed in 54\% of the \ce{In3Ni2}(001) configurations, which results in $W_{\mathrm{Cu}} = 0.83$~eV.

The material space also contains the main-group metals Ga, In, and K, which occur in 1,312 of the 2,613 facets.
Compared with the pure transition metals, the binary alloys of a transition metal with Ga or In bind *CO more weakly on average in all 21 such pairs of the dataset (median shift of the mean *CO adsorption energy of 0.49~eV) and show a larger fraction of desorbed *CO in 20 of the 21 pairs.
In contrast, the minimum *CO adsorption energy shifts by only 0.05~eV in the median, and the minimum adsorption energies of *OCHO and *H shift by less than 0.1~eV.
Alloying with Ga or In therefore mainly reduces the fraction of sites that bind *CO strongly rather than the binding strength of the strongest sites.
Such an analysis would not be possible with a descriptor based on the strongest site only, whereas the AED captures this change through the shape of the distribution and the desorption bin.
As a result, facets containing main-group metals make up half of the dataset but only 5 of the 53 facets within the inner threshold of Zn@Cu(211), including \ce{GaCu3}(110), GaPd(210), and \ce{K(InAu2)2}(101), and 3 of the 28 facets within the inner threshold of Ni(110).

Furthermore, as for any reference-based screening, the choice of the reference facet inherently influences the screening outcomes. 
Within the inner threshold, 51 of the 53 facets around Zn@Cu(211) contain Cu or Zn, which reflects a bias of the reference-based search toward compositions that resemble the reference.
Similarly, Ni-containing facets dominate the inner threshold of Ni(110), with 21 of the 28 facets.
Both biases are less prominent in the extended neighborhoods, in which 63 of the 203 additional facets around Zn@Cu(211) contain neither Cu nor Zn and 54 of the 113 additional facets around Ni(110) contain no Ni.
Still, additional reference facets, for example of ternary alloys, would further extend the accessible compositional space and allow the advantages of higher-order compositional complexity to be explored.

To obtain more insight into the ternary alloys, we tested whether their AEDs can be reproduced by mixtures of the AEDs of the facets of their constituent metals and binary alloys.
Most ternary facets are interpolations of their constituents, i.e., their AEDs are closely reproduced by such a mixture (Figure~\SIref{fig:ternary_mixture} of the SI), which agrees with the additive effect of the individual elements on the adsorption energies reported for high-entropy alloys~\cite{tan_ai_2025}.
A ternary phase nevertheless provides such a combination of adsorption sites in a single stable phase, which is more accessible to synthesis than a controlled mixture of several phases and facets.
\ce{K(InAu2)2}(101) is an exception, since no mixture of the facets of K, In, Au, and their binary alloys reaches its similarity to Zn@Cu(211).
On the In--Au binaries, *CO desorbs from a large fraction of the adsorption sites, whereas the K--Au binaries retain *CO but bind the O-bound intermediates less similarly to the reference.
The incorporation of K into In--Au alloys may therefore stabilize the adsorption of *CO and increase the activity toward methanol.

The overall results demonstrate the descriptive power of facet-resolved AEDs, despite their limitations.
The AED-similarity to Zn@Cu(211) first recovers Cu-based facets and subsequently identifies PdZn and GaPd facets, which have been reported for methanol synthesis, whereas the similarity to Ni(110) identifies Ni-, Co-, Ru-, and Rh-based facets known for methanation~\cite{musab_ahmed_ni-based_2024, godoy-gutierrez_co2_2025, guo_ru_2018, martin_structure_2018}.
These and previous examples show the strength of AEDs for finding energetically similar facets and their ability to pinpoint exceptional adsorption behavior.

\subsection{Facet- and Composition-Dependent Selectivity}

Using PC transformation on AED moments, we reduce the dimensionality of our descriptor space and obtain a physically interpretable map of the adsorption landscapes across the dataset.
The loading vectors and their contributions (see Figure~\SIref{fig:relative_contri_pca}) indicate that PC1 is dominated by moments of the O-bound intermediates (*\ce{OCH3}, *\ce{OCHO}, *\ce{OH}), with a small positive contribution from *CO.
On the other hand, PC2 is primarily governed by *\ce{CO} moments with minor contributions from the O-bound intermediates in the negative direction. 

As *H does not appear in the top 10 loading vectors of the PCA, we performed sensitivity analysis to assess whether excluding its adsorption-energy descriptors affects the embedding.
The corresponding PCA embedding is shown in Figure~\SIref{fig_sensitivity_analysis} of the SI.
The overall distribution of facets and organization of the latent space are not substantially modified relative to the PCA obtained using the moments of all adsorbates.

This AED-PCA map (Figure~\ref{fig:pca}) can be viewed as a two-dimensional ``beyond-Sabatier" plot, where the PC1 axis reflects variations in binding strengths of the O-bound intermediates and the PC2 axis captures systematic changes in *\ce{CO} binding. 
The separation of the two references along PC2 reflects the role of *CO binding in the transition between methanol formation and methanation.
Along PC1, at similar *CO binding, the stability of the O-bound intermediates changes, which affects the balance between methanol and less-hydrogenated products such as formic acid.
Both axes can be related to established trends in the reactivity of metal surfaces.
The *CO binding along PC2 weakens from the materials containing Fe, Co, Ni, Ru, Rh, or Ir through those containing Pd or Pt to the Cu-, Ag-, and Au-based materials and the \textit{sp} metals (Figure~\SIref{fig:aed_groups} of the SI), which reflects the filling of the \textit{d} band described by the \textit{d}-band model~\cite{norskov_density_2011, jones_using_2008}.
The binding of the O-bound intermediates along PC1 instead follows the oxophilicity of the metals~\cite{kepp_quantitative_2016} and is strongest for the materials containing early transition metals, which bind *CO only moderately.
A single electronic descriptor, such as the \textit{d}-band center, shifts the binding of all intermediates in the same direction and therefore cannot resolve these two independent directions, which the AED moments capture directly.

The compositions found around both references are in agreement with literature observations.
Around Zn@Cu(211), the closest facets belong to Cu--Zn, Cu--Ga, and Cu--Au materials, which agrees with the promotion of Cu-based methanol catalysts by Zn~\cite{kidd_methanol_1969, hausberger_catalyst_1981}, Ga~\cite{toyir_catalytic_2001, medina_catalytic_2017, lee_unraveling_2023, hansen_universal_2025}, and Au~\cite{xie_boosting_2023, pasupulety_studies_2015, mosrati_low-temperature_2023}.
The inner threshold also contains facets of the PdZn intermetallic (mp-1652, denoted ZnPd in the Materials Project and in the figures),  as well as GaPd(210). Both of these materials have been reported as methanol catalysts~\cite{brix_tuning_2020, lawes_co2_2024}.

Around Ni(110), the facets within the inner threshold mainly belong to Ni and to alloys of Ni with Zn, Pt, Fe, Co, Ga, and Mn, and both Ni-based catalysts~\cite{refaat_efficient_2023, zhi_comparative_2017, musab_ahmed_ni-based_2024} and Ni--Co alloys~\cite{deng_unsupported_2022, godoy-gutierrez_co2_2025} are known for their methanation activity.
Among the Ni-free facets in Table~\ref{tab:selectivity_map}, the Ru and Co facets agree with the methanation activity of Ru-based~\cite{guo_ru_2018, chai_boosting_2019} and Co-based catalysts~\cite{godoy-gutierrez_co2_2025, schubert_highly_2016}, and CoPt(211) relates to the Pt-promoted Co catalysts reported by Schubert et al.~\cite{schubert_highly_2016}.
Table~\ref{tab:selectivity_map} additionally contains Rh(100), which matches the methanation activity of supported Rh catalysts~\cite{martin_structure_2018, gomez-flores_modulating_2024}, and two FePt facets, which have been studied for \ce{CO2} hydrogenation~\cite{konopatsky_co2_2026}.
However, Fe-containing facets in the neighborhood of Ni(110) should be interpreted with caution, since Fe-based catalysts are more prominently known for C--C coupling~\cite{yang_recent_2025, yu_advances_2026}, which was outside the scope of this study.

While these experimental studies report the catalytic performance of the materials, the exact Miller planes found in the neighborhoods of both references may not be directly observed or characterized.
It is crucial to note that the catalytic surfaces are transient under reaction conditions and should be characterized \textit{in operando}.
The elevated temperature during the reaction can also lead to highly dynamic behavior of the catalytic nanoparticles, potentially exposing numerous facets, which makes our analysis particularly relevant.
Several facets in Table~\ref{tab:selectivity_map} belong to materials that, to our knowledge, have not been reported for \ce{CO2} hydrogenation and are therefore novel candidates for experimental assessment.
For methanol, these are the K--Au(--In) facets, i.e., four \ce{K(InAu2)2} facets led by \ce{K(InAu2)2}(101), \ce{KAu5}(102), \ce{KAu2}(101), and two \ce{K3InAu5} facets, together with \ce{YGa2Pd}(010) and \ce{Zn11Rh2}(110).
For methane, these are VIr(010), \ce{YPt3}(111), and \ce{MnGaCo2}(111).

\begin{figure}
    \centering
    \includegraphics[width=\linewidth]{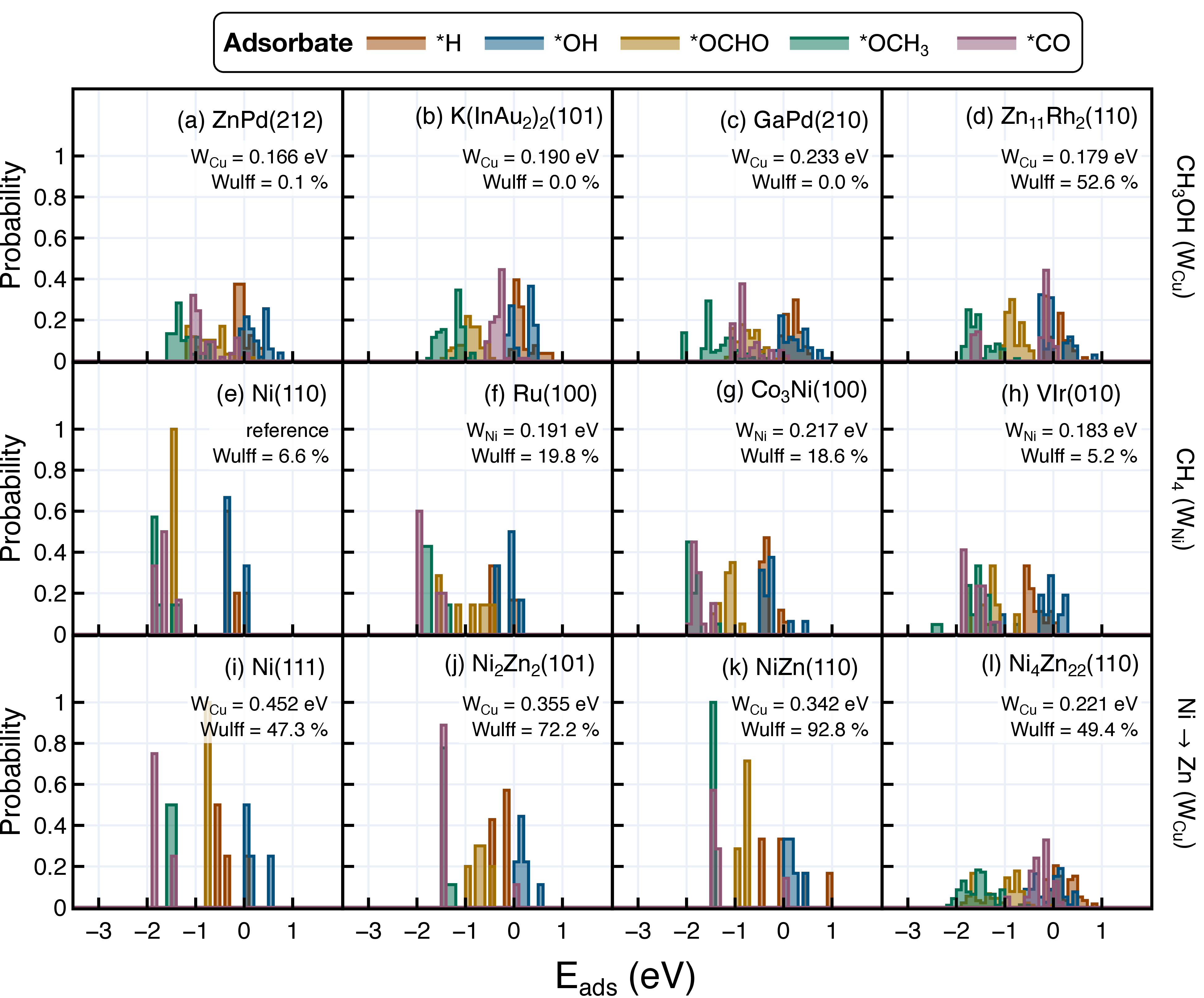}
    \caption{AEDs of example facets.
    (a--d) Facets within $W_{\mathrm{Cu}} \leq 0.25$~eV of Zn@Cu(211) that do not contain Cu: ZnPd(212), the closest such facet, \ce{K(InAu2)2}(101) and GaPd(210), the only such facets that contain neither Cu nor Zn, and \ce{Zn11Rh2}(110), the most abundant facet of the Zn-rich intermetallic \ce{Zn11Rh2}. The AED of Zn@Cu(211) is shown in Figure~\ref{fig:Zn@Cu(211)}(c).
    (e--h) The Ni(110) reference and facets within $W_{\mathrm{Ni}} \leq 0.25$~eV: Ru(100) and \ce{Co3Ni}(100), the most abundant facets of Ru and \ce{Co3Ni} within this threshold, and VIr(010), the closest facet of a material without Ni.
    (i--l) The most abundant facet of each material along the Ni--Zn series: Ni(111), \ce{Ni2Zn2}(101), NiZn(110), and \ce{Ni4Zn22}(110), annotated with $W_{\mathrm{Cu}}$.
    Each panel also gives the Wulff percentage of the facet.
    }
    \label{fig:top12_aeds}
\end{figure}

\begin{figure}
    \centering
    \includegraphics[width=0.6\linewidth]{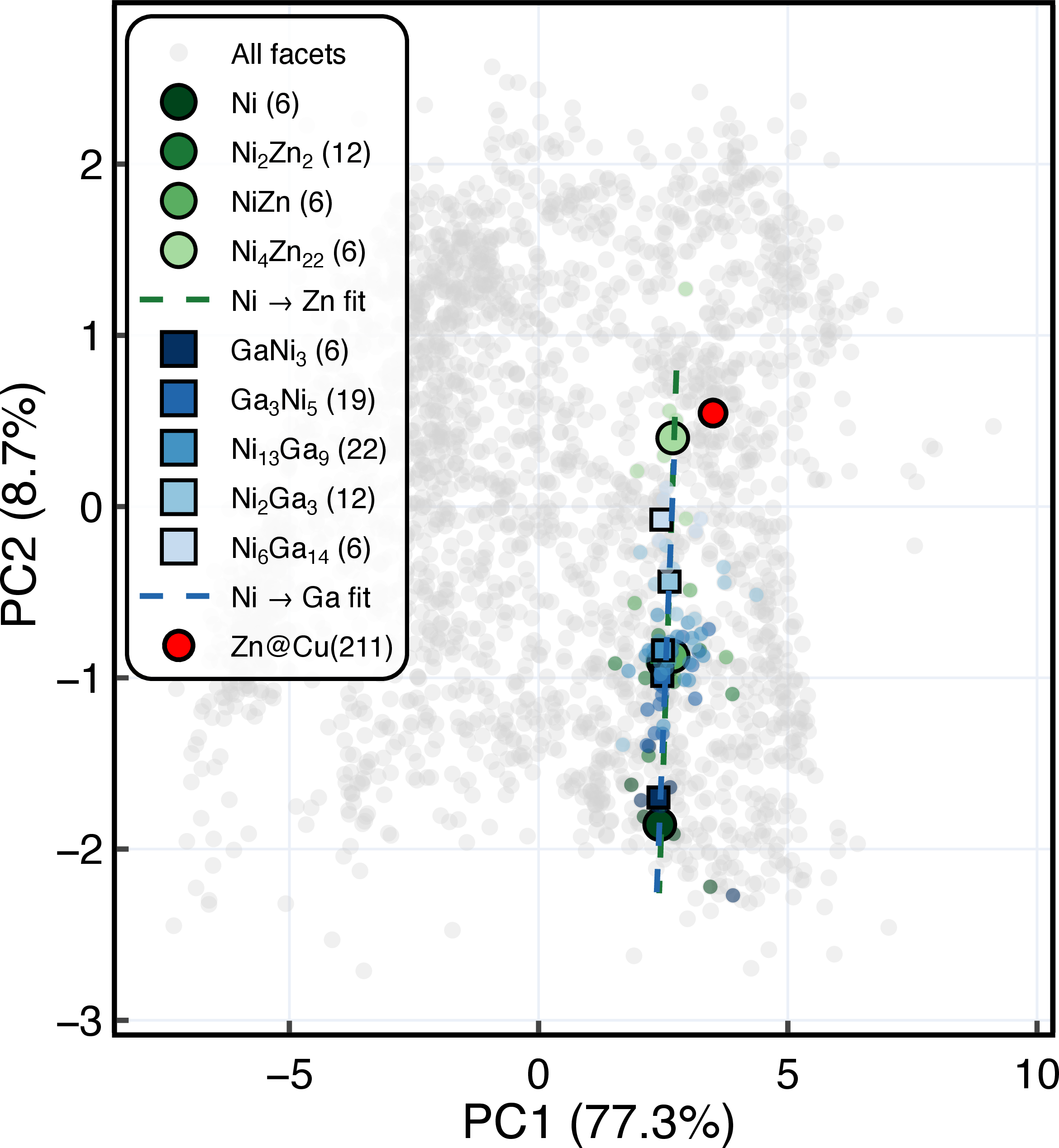}
    \caption{Facets of the Ni--Zn and Ni--Ga series in the PCA space of the six selected moments. Small markers show the individual facets of Ni, \ce{Ni2Zn2}, NiZn, and \ce{Ni4Zn22} (green) and of \ce{GaNi3}, \ce{Ga3Ni5}, \ce{Ni13Ga9}, \ce{Ni2Ga3}, and \ce{Ni6Ga14} (blue), and large circles (Ni--Zn) and squares (Ni--Ga) show the median PC1 and PC2 of each material, with the number of facets given in parentheses. The dashed lines are total-least-squares fits through the medians of each series, both starting from Ni. All other facets are shown in gray, and the red marker shows the reference Zn@Cu(211).}
    \label{fig:ni_series}
\end{figure}

Figure~\ref{fig:top12_aeds} compares the AEDs of example facets from the inner neighborhoods of both references and along the Ni--Zn series, which further clarifies the adsorption energy landscape.
For materials with several facets within the inner threshold, the figure shows the facet with the largest Wulff percentage, which is most likely to be exposed on the nanoparticle, except for ZnPd(212), which is shown as the closest facet without Cu.
For the facets in the top row (Figure~\ref{fig:top12_aeds}(a--d)), the median *\ce{CO} adsorption energies range from $-0.92$ to $-0.18$~eV, close to the median of $-0.51$~eV on Zn@Cu(211), and the minimum *\ce{OCHO} adsorption energies range from $-1.41$ to $-1.04$~eV, compared with $-1.27$~eV on the reference.
The facets in the middle row (Figure~\ref{fig:top12_aeds}(e--h)) bind *\ce{CO} considerably stronger, with median *\ce{CO} adsorption energies between $-1.92$ and $-1.55$~eV.
The same facets also bind *H stronger, with median *H adsorption energies between $-0.43$ and $-0.33$~eV, compared with values between $-0.02$ and 0.21~eV for the facets in the top row.
The stronger binding of *CO and *H on the facets around Ni(110) matches the methanation activity reported for Ni-, Co-, and Ru-based catalysts.
The bottom row (Figure~\ref{fig:top12_aeds}(i--l)) follows the Ni--Zn series, which alloys Ni, the metal of the methanation reference, with Zn, the promoter of the methanol reference.
Along the facets Ni(111), \ce{Ni2Zn2}(101), NiZn(110), and \ce{Ni4Zn22}(110), $W_{\mathrm{Cu}}$ decreases from 0.452~eV to 0.355, 0.342, and 0.221~eV, respectively, so that \ce{Ni4Zn22}(110) lies within the inner threshold of Zn@Cu(211).
With increasing Zn content, the median *CO adsorption energy gradually shifts from $-1.81$~eV on Ni(111) to $-0.37$~eV on \ce{Ni4Zn22}(110), and the median *H adsorption energy shifts from $-0.44$ to 0.15~eV.
This gradual transition is also visualized in Figure~\ref{fig:ni_series}.
Increasing Zn content therefore moves the Ni-based facets continuously from the neighborhood of Ni(110) toward the neighborhood of Zn@Cu(211), mainly along PC2.
With increasing Zn content, also the median $W_{\mathrm{Ni}}$ increases from 0.245~eV for Ni to 0.322 and 0.312~eV for NiZn and \ce{Ni2Zn2}, respectively, and to 0.487~eV for \ce{Ni4Zn22}.

Similarly illustrated in Figure~\ref{fig:ni_series}, alloying Ni with Ga also moves the facets away from Ni(110) along PC2.
Along the series Ni, \ce{GaNi3}, \ce{Ga3Ni5}, \ce{Ni13Ga9}, \ce{Ni2Ga3}, and \ce{Ni6Ga14}, the median PC2 increases from $-1.86$ to $-0.11$. Simultaneously, the median $W_{\mathrm{Ni}}$ increases from 0.245~eV for Ni to 0.735~eV for \ce{Ni6Ga14}, whereas the median $W_{\mathrm{Cu}}$ remains between 0.45 and 0.56~eV.
\ce{Ga3Ni5}, which corresponds to the \ce{Ni5Ga3} catalyst identified computationally.
The potential of  the \ce{Ni5Ga3} catalyst has been confirmed experimentally for methanol synthesis by Studt et al.~\cite{studt_discovery_2014} and it lies at similar median distances to both references, with $W_{\mathrm{Cu}} = 0.482$~eV and $W_{\mathrm{Ni}} = 0.504$~eV.
Zimmerli et al.~\cite{zimmerli_structure_2024} reported that alloying Ni with Ga suppresses methanation, with the highest methanol selectivity for a Ni:Ga ratio of approximately 75:25, and that \ce{GaO_x} species stabilize the formate intermediate.
The present descriptor space contains only metallic facets and therefore does not capture the role of \ce{GaO_x}, which may explain why the Ni--Ga facets move away from the methanation reference without reaching the inner neighborhood of Zn@Cu(211).
Similar composition-driven changes in selectivity, which originate from the modified electronic properties of the alloyed surfaces~\cite{wang_co2_2021}, have been reported experimentally, for example for the promotion of Ni by In~\cite{zhu_ni-synergy_2021} and for the switch of Pt from the RWGS reaction to methanol synthesis upon alloying with Ga~\cite{zhou_reactivity_2024}.

Formic acid formation is not assessed with the present approach, as heterogeneous HCOOH formation typically proceeds in the liquid phase in the presence of a base~\cite{sun_heterogeneous_2021}, rather than through thermal catalytic conversion.
Experimental studies on Pd catalysts modified by Ga oxide~\cite{mori_interplay_2023} and on Pd/ZnO~\cite{zhang_zinc_2020} report enhanced HCOOH formation, and the role of the composition of Pd-based catalysts in steering the product selectivity has been reviewed by Cao et al.~\cite{cao_selective_2025}. However, no facet-level reference catalyst for HCOOH formation has been established in the literature.
A facet-level reference for HCOOH formation, once established, could be used in the same way as Zn@Cu(211) and Ni(110).

Facets with weaker binding of both *CO and the O-bound intermediates, such as Ag and Au facets at large PC1 values, are expected to favor CO desorption and the RWGS pathway~\cite{ye_design_2025}.
Ag-based materials show low \ce{CO2} conversion and high RWGS selectivity~\cite{choi_catalytic_2017, zhang_blocking_2023, wei_pressure-controlled_2026}.
However, the RWGS reaction is endothermic and is therefore favored at higher temperatures than methanol synthesis~\cite{choi_catalytic_2017}, so that the selectivity toward CO depends strongly on the reaction temperature, which the adsorption energies alone do not capture.
In addition, a facet-level reference for selective CO formation over metallic catalysts is not established.
We therefore do not define a separate region for CO formation in the PCA space.

The neighborhoods of Zn@Cu(211) and Ni(110) in the PCA space (Figure~\ref{fig:pca}) reflect gradual boundaries rather than strict thresholds in terms of adsorption energies.
To translate the two neighborhoods into approximate adsorption energy ranges, we applied the inverse transformation of the first two principal components to the region within the inner threshold of each reference and recovered the corresponding minimum adsorption energies (Figure~\SIref{fig:ads_energy_ranges}).
The \ce{CH4} window binds *CO and *OCHO more strongly than the \ce{CH3OH} window, with minimum *CO adsorption energies of $-2.24$ and $-1.09$~eV, respectively, whereas the *\ce{OCH3} minima are similar and *H and *OH bind more weakly in the \ce{CH3OH} window.
The moderate binding of *CO and *OH in the \ce{CH3OH} window agrees with the analysis of Tang et al.~\cite{tang_electricity_2020}, although that analysis addressed electrochemical \ce{CO2} reduction.
Because the AEDs of both reference facets are narrower than those of a typical facet, the two-component reconstruction places the minima 0.02--0.42~eV below the actual minima of the reference facets, but the ordering between the two windows is retained.

To demonstrate how our workflow can be applied to materials beyond the screening domain, we performed a case study on an experimentally reported \ce{Cu4Pd} alloy, which lies slightly above the convex hull (0.016 eV per atom) yet is known to form under reaction conditions (Section~\SIref{sec:cu4pd_casestudy}).
All twelve \ce{Cu4Pd} facets lie within $W_{\mathrm{Cu}} \leq 0.40$~eV, and ten facets lie within the inner threshold, whereas the (212) facet, which dominates the equilibrium Wulff shape, lies at $W_{\mathrm{Cu}} = 0.274$~eV.
In the PCA space, the \ce{Cu4Pd} facets lie close to Zn@Cu(211), with slightly weaker binding of the O-bound intermediates and stronger *CO binding than the reference.
The \ce{Cu4Pd} facets therefore have a methanol-like adsorption landscape, and \ce{Cu4Pd} warrants experimental assessment for \ce{CO2} hydrogenation to methanol, although the Wulff-dominant (212) facet deviates most from the reference.

\subsection{Future Directions and Experimental Outlook}

The AED framework offers a descriptor-based perspective on activity and C1 product selectivity and could be extended to a broader chemical space, including the oxide components crucial to \ce{CO2} catalysis, once MLFFs of sufficient accuracy for oxides are available.
A generalized workflow would accelerate high-throughput screening, although several challenges remain for fully rational catalyst design.

Facets located near one another in the AED and latent spaces exhibit similar adsorption-energy characteristics and can therefore be prioritized as candidates with potentially related catalytic behavior.
However, these relationships should be interpreted in light of the approximations underlying the computed adsorption energies.
In particular, the MLFFs used in this work are spin-unpolarized, whereas spin effects can alter adsorption energies, especially for magnetic materials~\cite{escano_another_2009,cipriano_surface_2026,biz_magnetism_2020}.
Additional spin-polarized DFT calculations for representative systems (Figure~\SIref{fig:sp-nsp} and Table~\SIref{tab:sp-nsp} of the SI) reveal larger discrepancies for ferromagnetic materials than for diamagnetic and paramagnetic systems.
Although most sampled adsorption sites remain within the adopted EMAE threshold of 0.25~eV, larger site-specific errors can influence the AEDs and reduce the reliability of predictions for ferromagnetic materials.
Spin-aware universal MLFFs would therefore improve the robustness of large-scale screening across catalyst spaces.
Moreover, the experimentally observed activities and selectivity trends may nevertheless differ due to effects outside the present descriptor space, including strong metal-support interactions~\cite{zhang_strong_2024} and metal-oxide interfacial sites~\cite{lempelto_exploring_2023}.

PCA projections of AED moments encode qualitative trends in adsorption energies but do not explicitly resolve reaction barriers, surface coverages, reaction rates, or product formation rates. 
Reaction-network and barrier-based modeling can provide a more direct description of activity and selectivity~\cite{polynski_construction_nodate, kreitz_quantifying_2021}.
However, their application remains challenging due to near-zero predicted formation rates, reliance on empirical parameters, uncertainty propagated from adsorption-energy errors, and substantial computational cost. 
Accordingly, the selectivity trends inferred from the AED-PCA map should be interpreted as hypotheses grounded in adsorption-energy trends, rather than quantitative predictions of turnover frequencies.
Targeted microkinetic simulations and reaction-barrier calculations for a subset of promising candidates could further test these hypotheses and prioritize systems for experimental validation.
For the same subset, electronic-structure analyses, such as \textit{d}-band centers and widths, crystal orbital Hamilton populations, or Bader charges, could link the trends in the AEDs to their electronic origin, which is a topic reserved for future examination.

The AEDs are computed for single adsorbates on clean, bulk-terminated surfaces.
Under reaction conditions, adsorbate coverage can change the relative surface energies and thus the exposed facets, and adsorbate-induced segregation and reconstruction can alter the composition and coordination of the active sites.
The reference catalyst itself illustrates this effect, since the Zn decoration of the Cu steps forms under reaction conditions and depends on them~\cite{behrens_active_2012, kuld_quantifying_2016}.
The ranking therefore identifies candidates whose clean surfaces resemble the references, and the stability of these surfaces under reaction conditions has to be assessed for each candidate.
For a short list of candidates, ab initio thermodynamics of the surface terminations as a function of the gas-phase chemical potentials, coverage-dependent adsorption energies, and reconstruction simulations can test whether the ranking holds under reaction conditions, which is not feasible for the thousands of facets of the present screening.

Facet abundance is another critical dimension that links descriptor predictions to experimental reality.
A recurring observation in our screening is that facets with AED fingerprints similar to the Zn@Cu(211) reference frequently exhibit low or zero theoretical abundance in the equilibrium Wulff construction (Figure~\ref{fig:wsd_vs_facets}).
Of the 53 facets within the inner threshold, 22 facets (42\%) do not appear on the equilibrium Wulff shape, compared with 32\% of all facets with a Wulff construction.
Wulff constructions based on MLFF surface energies estimate facet populations for idealized single-crystal equilibrium morphologies. Although these results indicate that certain terminations are thermodynamically disfavored within the vacuum Wulff model, they do not preclude their formation or catalytic relevance under realistic synthesis and reaction conditions~\cite{ringe_kinetic_2013, xing_temperature-dependent_2020}.

The occurrence of low-abundance facets near the Zn@Cu(211) reference, therefore, points to the need for synthesis strategies that selectively access and retain these terminations. 
Facet-engineering approaches~\cite{cheng_facet-engineering_2022, shi_engineering_2022}, kinetically controlled growth~\cite{xiao_high-index-facet-_2020}, and tailored metal-support interactions~\cite{maxson_metal-support_2025} may provide routes to stabilize surfaces that are not predicted to dominate the vacuum equilibrium morphology.
Experimental validation should consequently assess whether the prioritized facet-alloy combinations can be synthesized.

Beyond controlling facet exposure, alloying and electronic modification offer complementary routes to improved catalytic performance by tuning adsorption energies across exposed surface sites.
In practice, the optimal design pathway depends on synthetic feasibility, long-term stability under reaction conditions, and the availability of supports or promoters that can realize the desired facet and alloy structures.
We therefore envisage future work that combines descriptor-guided selection of facet-alloy combinations with targeted synthesis, microkinetic and barrier modeling, and operando characterization to build a more quantitative bridge between AED-based maps and experimentally measured activity and selectivity.

%%%%%%%%%%%%%%%%%%%%%%%%%%%%%%%%%%%%%%%%%%%%%%%%%%%%%%%%%%%%%%%%%%%%%%%%%%%%%%%%%%%%%%%%%%%%
\section{Conclusions}
%%%%%%%%%%%%%%%%%%%%%%%%%%%%%%%%%%%%%%%%%%%%%%%%%%%%%%%%%%%%%%%%%%%%%%%%%%%%%%%%%%%%%%%%%%%%
In this work, we used MLFF-driven high-throughput simulations to construct facet-resolved AEDs for metallic, binary, and ternary alloy catalysts for thermal \ce{CO2} hydrogenation. 
By computing facet-resolved AED descriptors using an accelerated workflow to scan a broad chemical space, our analysis captures the underlying energy landscape beyond single-site descriptors and provides information on both activity and selectivity.
Distribution-level similarity in AED space provides a practical proxy for apparent activity, while projection of AED moments onto a low-dimensional latent space reveals systematic trends in the binding of *CO and of the O-bound intermediates, which are associated with the C1 product selectivity, across different facets and alloy compositions. 
Using the experimentally motivated Zn-doped Cu(211) and Ni(110) as reference facets for methanol and methane formation, 
the similarity analysis pinpoints the methanol- and methane-selective areas in the low-dimensional latent space. It also allows us to identify compositionally novel candidates such as \ce{K(InAu2)2}(101) for methanol and VIr(010) for methane.
The resulting workflow yields a prioritized list of composition-facet combinations for experimental validation in \ce{CO2} hydrogenation.
Overall, this framework provides a route toward rational, data-driven catalyst discovery by linking broad chemical-space exploration with interpretable representations and statistical learning of the adsorption energy landscape.

\begin{acknowledgement}

The authors would like to thank Annukka Santasalo-Aarnio and Arpad Toldy for fruitful discussions. 
O.K. and P.P. express their gratitude to Kirby Broderick, Adeesh Koluru, Brook Wander, John Kitchin, and other Kitchin Research Group members at Carnegie Mellon University, as well as Zachary Ulissi at Meta, for their assistance with the OCP models.
This project received funding from the Deutsche Forschungsgemeinschaft (DFG, German Research Foundation) under Germany's Excellence Strategy - EXC 2089/2 - 390776260, and the European Union - NextGenerationEU instrument from the Research Council of Finland's AICon project (grant number 348179).
O.K. was supported by the Research Council of Finland under project number 371666.
The authors further gratefully acknowledge CSC - IT Center for Science, Finland, and the Aalto Science-IT project for generous computational resources.

The authors used LLM tools to strengthen the language of this manuscript. The authors reviewed and edited the content and take full responsibility for the content of the publication.

\end{acknowledgement}

\section{Data Availability}

All data are available on Zenodo at DOI: \texttt{10.5281/zenodo.24199962}~\cite{pisal_dataset_2026}.

\begin{suppinfo}
Slab Thickness Tests;
Adsorption energies of *H on VPt surfaces calculated using the \texttt{equiformerV2} MLFF on 7 Å and 14 Å slab models;
Comparison of *H adsorption energies on 7~\AA{} and 14~\AA{} slabs for sampled surface configurations of VPt alloy calculated using DFT;
Statistics on adsorption energy prediction;
All 226 materials used in the final evaluation;
Materials removed from the final evaluation;
Anomalous adsorbate geometries;
AEDs of example facets with anomalous configurations;
AEDs of the facets closest to Zn@Cu(211), excluding Cu-containing materials;
AEDs of the facets closest to Zn@Cu(211), excluding Cu- and Zn-containing materials;
AEDs of the facets closest to Ni(110), excluding Ni-containing materials;
Group-mean AEDs of all adsorbates for the chemistry groups of the 226 materials;
Selection of statistical moments for the PCA;
Loading values of the top 10 moments for the first two principal components;
Ternary facets as mixtures of the facets of their constituent metals and binary alloys;
Sensitivity analysis to assess the contribution of *H AED moments in the latent space of 2D PCA;
Minimum adsorption energy ranges of the \ce{CH3OH} and \ce{CH4} windows obtained from the inverse PCA transform;
Facets with $W_{\mathrm{Cu}} \leq 0.40$~eV to Zn@Cu(211);
Facets with $W_{\mathrm{Ni}} \leq 0.40$~eV to Ni(110);
Parity plot comparing adsorption energies calculated with spin polarization (SP) and without spin polarization (NSP) for sampled adsorbate--slab configurations of CoPt$_3$, Ni, Cu, and V surfaces;
Mean absolute error (MAE) between spin-polarized (SP) and non-spin-polarized (NSP) adsorption energies for the four studied materials;
Case Study: \ce{Cu4Pd} (mp-30594); Projection of the facet-resolved AED moments of the out-of-domain \ce{Cu4Pd} alloy into the PCA space constructed for the screened material set;
Equilibrium Wulff percentages and distances $W_{\mathrm{Cu}}$ of the facet-resolved AEDs for the out-of-domain \ce{Cu4Pd} alloy.

\end{suppinfo}

\bibliography{achemso-demo}
% \printbibliography
\end{document}